\documentclass[prd,amsmath,aps,floats,amssymb, floatfix,superscriptaddress,nofootinbib,twocolumn,preprintnumbers]{revtex4-2}

\setcitestyle{authoryear,round}

\usepackage[T1]{fontenc}
\usepackage{aecompl}

\usepackage{newtxtext,newtxmath}
\usepackage{mdframed,bm}
\usepackage[T1]{fontenc}
\usepackage{ae,aecompl}
\usepackage{graphicx}	
\usepackage{amsmath}	
\usepackage{amssymb}	
\usepackage{booktabs}
\usepackage{color}
\usepackage{enumitem}
\usepackage[colorlinks=true]{hyperref}
\hypersetup{
    urlcolor=blue,
    citecolor=blue,
    linkcolor=blue,
  }

\makeatletter
  \let\NAT@sort\z@
  \let\NAT@cmprs\z@
\makeatother

\usepackage{graphicx}
\usepackage{dcolumn}
\usepackage{bm}

\preprint{APS/123-QED}

\usepackage{amsmath}
\usepackage{fontawesome5}
\usepackage{color}
\usepackage{xcolor}
\usepackage{ulem}

\usepackage{graphicx}	
\usepackage{amsmath}	

\usepackage{fontawesome5}
\usepackage{color}
\usepackage{xcolor}
\usepackage{xspace}

\usepackage{enumitem}
\setlist{nosep}

\newcommand{\eg}{{\sl e.g.}, }

\newcommand{\msun}{\ensuremath{\, {\rm M}_\odot}} 
         
\newcommand{\mpc}{\ensuremath{\, {\rm Mpc}}}

\newcommand{\chisq}{\chi^2}

\newcommand\arcdeg{\mbox{$^\circ$}}%

\renewcommand{\eg}{{\sl e.g.}, }   
\newcommand{\ie}{{\sl i.e.}, }    
\newcommand{\Om}{\Omega_{\rm m}}
\newcommand{\Seight}{S_8}
\newcommand{\LCDM}{$\Lambda$CDM\xspace}
\newcommand{\wCDM}{$w$CDM\xspace}

\newcommand{\decade}{\textsc{DECADE}\xspace}

\newcommand{\HMx}{\textsc{HMx}\xspace}
\newcommand{\Hmx}{\HMx}
\newcommand{\BCEmu}{\textsc{BCEmu}\xspace}
\newcommand{\Bacco}{\textsc{Bacco}\xspace}
\newcommand{\BaryonForge}{\textsc{BaryonForge}\xspace}

\definecolor{orcidlogocol}{HTML}{A6CE39}
\definecolor{purple}{RGB}{128, 0, 128}
\definecolor{kelly}{RGB}{76, 187, 23}

\newcommand{\OrcidID}[1]{ \href[urlcolor = red]{https://orcid.org/#1}{\textcolor{lightgray}{\faOrcid}}}
\newcommand{\OrcidIDName}[2]{\href{https://orcid.org/#1}{#2}}

\defcitealias{y3-shapecatalog}{\textsc{GS21}}
\defcitealias{paper1}{\textsc{Paper I}}
\defcitealias{paper2}{\textsc{Paper II}}
\defcitealias{paper3}{\textsc{Paper III}}
\defcitealias{paper4}{\textsc{Paper IV}}

\newcommand*{\vcenteredhbox}[1]{\begingroup
\setbox0=\hbox{#1}\parbox{\wd0}{\box0}\endgroup}

\begin{document}

\title{\Large The Dark Energy Camera All Data Everywhere cosmic shear project V: Constraints on cosmology and astrophysics from 270 million galaxies\\ across 13,000 deg$^2$ of the sky}

\author{\OrcidIDName{0000-0002-6021-8760}{D.~Anbajagane} (\vcenteredhbox{\includegraphics[height=1.2\fontcharht\font`\B]{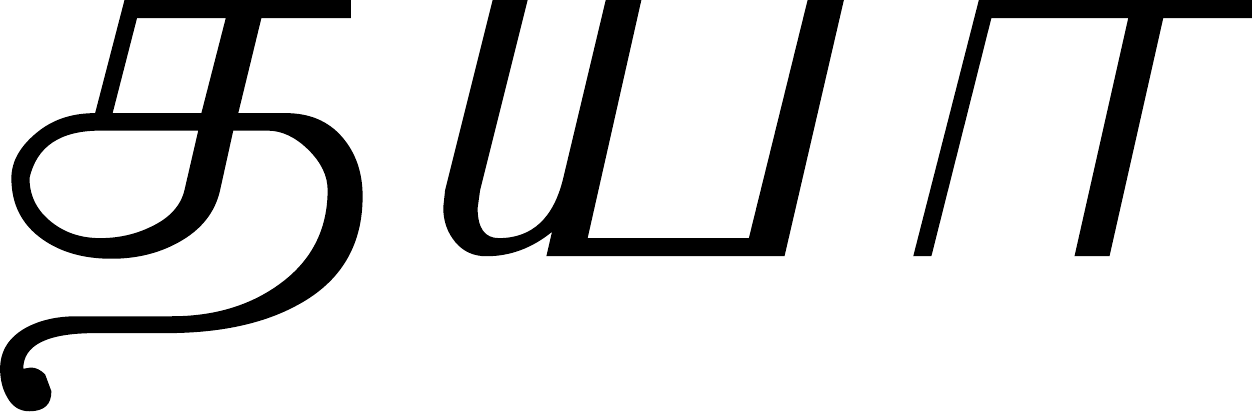}})$^\star$}
\affiliation{Department of Astronomy and Astrophysics, University of Chicago, Chicago, IL 60637, USA}
\affiliation{Kavli Institute for Cosmological Physics, University of Chicago, Chicago, IL 60637, USA}
\email{dhayaa@uchicago.edu, chihway@kicp.uchicago.edu}
\affiliation{NSF-Simons AI Institute for the Sky (SkAI), 172 E. Chestnut St., Chicago, IL 60611, USA}

\author{\OrcidIDName{0000-0002-7887-0896}{C.~Chang}$^\star$}
\affiliation{Department of Astronomy and Astrophysics, University of Chicago, Chicago, IL 60637, USA}
\affiliation{Kavli Institute for Cosmological Physics, University of Chicago, Chicago, IL 60637, USA}
\affiliation{NSF-Simons AI Institute for the Sky (SkAI), 172 E. Chestnut St., Chicago, IL 60611, USA}

\author{\OrcidIDName{0000-0001-8251-933X}{A.~Drlica-Wagner}}
\affiliation{Fermi National Accelerator Laboratory, P. O. Box 500, Batavia, IL 60510, USA}
\affiliation{Department of Astronomy and Astrophysics, University of Chicago, Chicago, IL 60637, USA}
\affiliation{Kavli Institute for Cosmological Physics, University of Chicago, Chicago, IL 60637, USA}
\affiliation{NSF-Simons AI Institute for the Sky (SkAI), 172 E. Chestnut St., Chicago, IL 60611, USA}

\author{\OrcidIDName{0000-0003-0478-0473}{C.~Y.~Tan}}
\affiliation{Department of Physics, University of Chicago, Chicago, IL 60637, USA}
\affiliation{Kavli Institute for Cosmological Physics, University of Chicago, Chicago, IL 60637, USA}

\author{\OrcidIDName{0000-0002-6904-359X}{M.~Adamow}}
\affiliation{Center for Astrophysical Surveys, National Center for Supercomputing Applications, 1205 West Clark St., Urbana, IL 61801, USA}
\affiliation{Department of Astronomy, University of Illinois at Urbana-Champaign, 1002 W. Green Street, Urbana, IL 61801, USA}

\author{\OrcidIDName{0000-0002-4588-6517}{R.~A.~Gruendl}}
\affiliation{Center for Astrophysical Surveys, National Center for Supercomputing Applications, 1205 West Clark St., Urbana, IL 61801, USA}
\affiliation{Department of Astronomy, University of Illinois at Urbana-Champaign, 1002 W. Green Street, Urbana, IL 61801, USA}

\author{\OrcidIDName{0000-0002-6002-4288}{L.~F.~Secco}}
\affiliation{Kavli Institute for Cosmological Physics, University of Chicago, Chicago, IL 60637, USA}

\author{\OrcidIDName{0000-0002-7523-582X}{Z.~Zhang}}
\affiliation{Department of Astronomy and Astrophysics, University of Chicago, Chicago, IL 60637, USA}
\affiliation{Department of Physics, Stanford University, 382 Via Pueblo Mall, Stanford, CA 94305, USA}
\affiliation{SLAC National Accelerator Laboratory, Menlo Park, CA 94025, USA}

\author{\OrcidIDName{0000-0001-7774-2246}{M.~R.~Becker}}
\affiliation{Argonne National Laboratory, 9700 South Cass Avenue, Lemont, IL 60439, USA}

\author{\OrcidIDName{0000-0001-6957-1627}{P.~S.~Ferguson}}
\affiliation{DIRAC Institute, Department of Astronomy, University of Washington, 3910 15th Ave NE, Seattle, WA, 98195, USA}

\author{\OrcidIDName{0009-0005-1143-495X}{N.~Chicoine}}
\affiliation{Department of Astronomy and Astrophysics, University of Chicago, Chicago, IL 60637, USA}
\affiliation{Department of Physics and Astronomy, University of Pittsburgh, 3941 O’Hara Street, Pittsburgh, PA 15260}

\author{\OrcidIDName{0000-0003-4394-7491}{K.~Herron}}
\affiliation{Department of Physics and Astronomy, Dartmouth College, Hanover, NH 03755, USA}

\author{\OrcidIDName{0000-0001-8505-1269}{A.~Alarcon}}
\affiliation{Institute of Space Sciences (ICE, CSIC),  Campus UAB, Carrer de Can Magrans, s/n,  08193 Barcelona, Spain}

\author{\OrcidIDName{0000-0002-5279-0230}{R.~Teixeira}}
\affiliation{Department of Astronomy and Astrophysics, University of Chicago, Chicago, IL 60637, USA}
\affiliation{Department of Physics, Duke University Durham, NC 27708, USA}

\author{\OrcidIDName{0000-0003-2911-2025}{D.~Suson}}
\affiliation{Department of Chemistry and Physics, Purdue University Northwest 2200, 169th Ave, Hammond, IN 46323}

\author{\OrcidIDName{0000-0001-7836-2261}{A.~J.~Shajib}}
\affiliation{Department of Astronomy and Astrophysics, University of Chicago, Chicago, IL 60637, USA}
\affiliation{Kavli Institute for Cosmological Physics, University of Chicago, Chicago, IL 60637, USA}
\affiliation{Center for Astronomy, Space Science and Astrophysics, Independent University, Bangladesh, Dhaka 1229, Bangladesh}

\author{\OrcidIDName{0000-0001-7836-2261}{J.~A.~Frieman}}
\affiliation{Department of Astronomy and Astrophysics, University of Chicago, Chicago, IL 60637, USA}
\affiliation{Kavli Institute for Cosmological Physics, University of Chicago, Chicago, IL 60637, USA}
\affiliation{SLAC National Accelerator Laboratory, Menlo Park, CA 94025, USA}
\affiliation{Kavli Institute for Particle Astrophysics \& Cosmology, P.O.\ Box 2450, Stanford University, Stanford, CA 94305, USA}

\author{\OrcidIDName{0000-0002-3173-2592}{A.~N.~Alsina}}
\affiliation{Instituto de Física Gleb Wataghin, Universidade Estadual de Campinas, 13083-859, Campinas, SP, Brazil}

\author{\OrcidIDName{0000-0002-6445-0559}{A.~Amon}}
\affiliation{Department of Astrophysical Sciences, Princeton University, Peyton Hall, Princeton, NJ 08544, USA}

\author{\OrcidIDName{0000-0003-0171-6900}{F.~Andrade-Oliveira}}
\affiliation{Physik-Institut, University of Zurich, Winterthurerstrasse 190, CH-8057 Zurich, Switzerland}

\author{\OrcidIDName{0000-0002-4687-4657}{J.~Blazek}}
\affiliation{Department of Physics, Northeastern University, Boston, MA 02115, USA}

\author{\OrcidIDName{0000-0003-4383-2969}{C.~R.~Bom}}
\affiliation{Centro Brasileiro de Pesquisas F\'isicas, Rua Dr. Xavier Sigaud 150, 22290-180 Rio de Janeiro, RJ, Brazil}

\author{\OrcidIDName{0000-0001-5871-0951}{H.~Camacho}}
\affiliation{Physics Department, Brookhaven National Laboratory, Upton, NY 11973}

\author{\OrcidIDName{0000-0002-3690-105X}{J.~A.~Carballo-Bello}}
\affiliation{Instituto de Alta Investigaci\'on, Universidad de Tarapac\'a, Casilla 7D, Arica, Chile}

\author{\OrcidIDName{0000-0003-3044-5150}{A.~Carnero~Rosell}}
\affiliation{Universidad de La Laguna, Dpto. Astrofísica, E-38206 La Laguna, Tenerife, Spain}
\affiliation{Instituto de Astrofisica de Canarias, E-38205 La Laguna, Tenerife, Spain}
\affiliation{Laborat\'orio Interinstitucional de e-Astronomia - LIneA, Rua Gal. Jos\'e Cristino 77, Rio de Janeiro, RJ - 20921-400, Brazil}

\author{\OrcidIDName{0000-0003-2965-6786}{R.~Cawthon}}
\affiliation{Physics Department, William Jewell College, Liberty, MO, 64068}

\author{\OrcidIDName{0000-0003-1697-7062}{W.~Cerny}}
\affiliation{Department of Astronomy, Yale University, New Haven, CT 06520, USA}

\author{\OrcidIDName{0000-0002-5636-233X}{A.~Choi}}
\affiliation{NASA Goddard Space Flight Center, 8800 Greenbelt Rd, Greenbelt, MD 20771, USA}

\author{\OrcidIDName{0000-0003-1680-1884}{Y.~Choi}}
\affiliation{NSF National Optical-Infrared Astronomy Research Laboratory, 950 North Cherry Avenue, Tucson, AZ 85719, USA}

\author{\OrcidIDName{0000-0002-8446-3859}{S.~Dodelson}}
\affiliation{Kavli Institute for Cosmological Physics, University of Chicago, Chicago, IL 60637, USA}
\affiliation{Fermi National Accelerator Laboratory, P. O. Box 500, Batavia, IL 60510, USA}
\affiliation{Department of Astronomy and Astrophysics, University of Chicago, Chicago, IL 60637, USA}

\author{\OrcidIDName{0000-0003-4480-0096}{C.~Doux}}
\affiliation{Université Grenoble Alpes, CNRS, LPSC-IN2P3, 38000 Grenoble, France}

\author{\OrcidIDName{0000-0002-1407-4700}{K.~Eckert}}
\affiliation{Department of Physics and Astronomy, University of Pennsylvania, Philadelphia, PA 19104, USA}

\author{\OrcidIDName{0000-0001-5148-9203}{J.~Elvin-Poole}}
\affiliation{Department of Physics and Astronomy, University of Waterloo, 200 University Ave W, Waterloo, ON N2L 3G1, Canada}

\author{\OrcidIDName{0000-0003-4373-2386}{J.~Esteves}}
\affiliation{Department of Physics, Harvard University, MA 02138, USA}

\author{\OrcidIDName{0000-0001-6134-8797}{M.~Gatti}}
\affiliation{Kavli Institute for Cosmological Physics, University of Chicago, Chicago, IL 60637, USA}

\author{\OrcidIDName{0000-0002-3730-1750}{G.~Giannini}}
\affiliation{Kavli Institute for Cosmological Physics, University of Chicago, Chicago, IL 60637, USA}

\author{\OrcidIDName{0000-0003-3270-7644}{D.~Gruen}}
\affiliation{University Observatory, Faculty of Physics, Ludwig-Maximilians-Universität, Scheinerstr. 1, 81679 Munich, Germany; Excellence Cluster ORIGINS, Boltzmannstr. 2, 85748 Garching, Germany}

\author{\OrcidIDName{0000-0001-9994-1115}{W.~G.~Hartley}}
\affiliation{Department of Astronomy, University of Geneva, ch. d’Ecogia 16, 1290 Versoix, Switzerland}

\author{\OrcidIDName{0000-0001-6718-2978}{K.~Herner}}
\affiliation{Fermi National Accelerator Laboratory, P. O. Box 500, Batavia, IL 60510, USA}

\author{\OrcidIDName{0000-0002-9378-3424}{E.~M.~Huff}}
\affiliation{Jet Propulsion Laboratory, California Institute of Technology, 4800 Oak Grove Dr., Pasadena, CA 91109, USA}

\author{\OrcidIDName{0000-0002-8220-3973}{B.~Jain}}
\affiliation{Department of Physics and Astronomy, University of Pennsylvania, Philadelphia, PA 19104, USA}

\author{\OrcidIDName{0000-0001-5160-4486}{D.~J.~James}}
\affiliation{Applied Materials Inc., 35 Dory Road, Gloucester, MA 01930}
\affiliation{ASTRAVEO LLC, PO Box 1668, Gloucester, MA 01931}

\author{\OrcidIDName{0000-0002-4179-5175}{M.~Jarvis}}
\affiliation{Department of Physics and Astronomy, University of Pennsylvania, Philadelphia, PA 19104, USA}

\author{\OrcidIDName{0000-0001-8356-2014}{E.~Krause}}
\affiliation{Department of Astronomy/Steward Observatory, University of Arizona, Tucson, AZ 85721 USA}

\author{\OrcidIDName{0000-0003-2511-0946}{N.~Kuropatkin}}
\affiliation{Fermi National Accelerator Laboratory, P. O. Box 500, Batavia, IL 60510, USA}

\author{\OrcidIDName{0000-0002-9144-7726}{C.~E.~Mart\'inez-V\'azquez}}
\affiliation{International Gemini Observatory/NSF NOIRLab, 670 N. A'ohoku Place, Hilo, Hawai'i, 96720, USA}

\author{\OrcidIDName{0000-0002-8093-7471}{P.~Massana}}
\affiliation{NSF's NOIRLab, Casilla 603, La Serena, Chile}

\author{\OrcidIDName{0000-0003-3519-4004}{S.~Mau}}
\affiliation{Department of Physics, Stanford University, 382 Via Pueblo Mall, Stanford, CA 94305, USA}
\affiliation{Kavli Institute for Particle Astrophysics \& Cosmology, P.O.\ Box 2450, Stanford University, Stanford, CA 94305, USA}

\author{\OrcidIDName{0000-0002-4475-3456}{J.~McCullough}}
\affiliation{Department of Astrophysical Sciences, Peyton Hall, Princeton University, Princeton, NJ USA 08544}

\author{\OrcidIDName{0000-0003-0105-9576}{G.~E.~Medina}}
\affiliation{Dunlap Institute for Astronomy \& Astrophysics, University of Toronto, 50 St George Street, Toronto, ON M5S 3H4, Canada}
\affiliation{David A. Dunlap Department of Astronomy \& Astrophysics, University of Toronto, 50 St George Street, Toronto ON M5S 3H4, Canada}

\author{\OrcidIDName{0000-0001-9649-4815}{B.~Mutlu-Pakdil}}
\affiliation{Department of Physics and Astronomy, Dartmouth College, Hanover, NH 03755, USA}

\author{\OrcidIDName{0000-0001-6145-5859}{J.~Myles}}
\affiliation{Department of Astrophysical Sciences, Princeton University, Peyton Hall, Princeton, NJ 08544, USA}

\author{\OrcidIDName{0000-0001-9438-5228}{M. ~ Navabi}}
\affiliation{Department of Physics, University of Surrey, Guildford GU2 7XH, UK}

\author{\OrcidIDName{0000-0002-8282-469X}{N.~E.~D.~Noël}}
\affiliation{Department of Physics, University of Surrey, Guildford GU2 7XH, UK}

\author{\OrcidIDName{0000-0002-6021-8760}{A.~B.~Pace}}
\affiliation{Department of Astronomy, University of Virginia, 530 McCormick Road, Charlottesville, VA 22904, USA}

\author{\OrcidIDName{000-0001-5780-637X}{S.~Pandey}}
\affiliation{Department of Physics, Columbia University, 538 West 120th Street, New York, NY, USA 10027, USA}
\affiliation{Columbia Astrophysics Laboratory, Columbia University, 550 West 120th Street, New York, NY 10027, USA}

\author{\OrcidIDName{0000-0002-2762-2024}{A.~Porredon}}
\affiliation{Centro de Investigaciones Energ\'eticas, Medioambientales y Tecnol\'ogicas (CIEMAT), Madrid, Spain}

\author{\OrcidIDName{0000-0002-5933-5150}{J.~Prat}}
\affiliation{Nordita, KTH Royal Institute of Technology and Stockholm University, SE-106 91 Stockholm.}

\author{\OrcidIDName{0000-0002-7354-3802}{M.~Raveri}}
\affiliation{Department of Physics and INFN, University of Genova, Genova, Italy}

\author{\OrcidIDName{0000-0001-5805-5766}{A.~H.~Riley}}
\affiliation{Institute for Computational Cosmology, Department of Physics, Durham University, South Road, Durham DH1 3LE, UK}

\author{\OrcidIDName{0000-0001-9376-3135}{E.~S.~Rykoff}}
\affiliation{SLAC National Accelerator Laboratory, Menlo Park, CA 94025, USA}
\affiliation{Kavli Institute for Particle Astrophysics \& Cosmology, P.O.\ Box 2450, Stanford University, Stanford, CA 94305, USA}

\author{\OrcidIDName{0000-0002-1594-1466}{J.~D.~Sakowska}}
\affiliation{Department of Physics, University of Surrey, Guildford GU2 7XH, UK}

\author{\OrcidIDName{0000-0001-7147-8843}{S.~Samuroff}}
\affiliation{Institut de F\'{i}sica d'Altes Energies, The Barcelona Institute of Science and Technology, Campus UAB, 08193 Bellaterra (Barcelona) Spain}

\author{\OrcidIDName{0000-0003-3054-7907}{D.~Sanchez-Cid}}
\affiliation{Centro de Investigaciones Energéticas, Medioambientales y Tecnológicas (CIEMAT), Madrid, Spain}
\affiliation{Physik-Institut, University of Zurich, Winterthurerstrasse 190, CH-8057 Zurich, Switzerland}

\author{\OrcidIDName{0000-0003-4102-380X}{D.~J.~Sand}}
\affiliation{Steward Observatory, University of Arizona, 933 North Cherry Avenue, Tucson, AZ 85721-0065, USA}

\author{\OrcidIDName{0000-0003-3402-6164}{L.~Santana-Silva}}
\affiliation{Centro Brasileiro de Pesquisas F\'isicas, Rua Dr. Xavier Sigaud 150, 22290-180 Rio de Janeiro, RJ, Brazil}

\author{\OrcidIDName{0000-0002-1831-1953}{I.~Sevilla-Noarbe}}
\affiliation{Centro de Investigaciones Energ\'eticas, Medioambientales y Tecnol\'ogicas (CIEMAT), Madrid, Spain}

\author{\OrcidIDName{0000-0002-6389-5409}{T.~Shin}}
\affiliation{Department of Physics, Carnegie Mellon University, Pittsburgh, PA 15213}

\author{\OrcidIDName{0000-0001-6082-8529}{M.~Soares-Santos}}
\affiliation{Physik-Institut, University of Zurich, Winterthurerstrasse 190, CH-8057 Zurich, Switzerland}

\author{\OrcidIDName{0000-0003-1479-3059}{G.~S.~Stringfellow}}
\affiliation{Center for Astrophysics and Space Astronomy, University of Colorado, 389 UCB, Boulder, CO 80309-0389, USA}

\author{\OrcidIDName{0000-0001-7836-2261}{C.~To}}
\affiliation{Kavli Institute for Cosmological Physics, University of Chicago, Chicago, IL 60637, USA}
\affiliation{NSF-Simons AI Institute for the Sky (SkAI), 172 E. Chestnut St., Chicago, IL 60611, USA}

\author{\OrcidIDName{0000-0002-9599-310X}{E.~J.~Tollerud}}
\affiliation{Space Telescope Science Institute, 3700 San Martin Drive, Baltimore, MD 21218, USA}

\author{\OrcidIDName{0009-0002-4207-0210}{A.~Tong}}
\affiliation{Department of Physics and Astronomy, University of Pennsylvania, Philadelphia, PA 19104, USA}

\author{\OrcidIDName{0000-0002-5622-5212}{M.~A.~Troxel}}
\affiliation{Department of Physics, Duke University Durham, NC 27708, USA}

\author{\OrcidIDName{0000-0003-4341-6172}{A.~K.~Vivas}}
\affiliation{Cerro Tololo Inter-American Observatory/NSF NOIRLab, Casilla 603, La Serena, Chile}

\author{\OrcidIDName{0000-0003-1585-997X}{M.~Yamamoto}}
\affiliation{Department of Astrophysical Sciences, Princeton University, Peyton Hall, Princeton, NJ 08544, USA}

\author{\OrcidIDName{0000-0002-9541-2678}{B.~Yanny}}
\affiliation{Fermi National Accelerator Laboratory, PO Box 500, Batavia, IL, 60510, USA}

\author{\OrcidIDName{0009-0006-5604-9980}{B.~Yin}}
\affiliation{Department of Physics, Duke University Durham, NC 27708, USA}

\author{\OrcidIDName{0000-0001-6455-9135}{A.~Zenteno}}
\affiliation{Cerro Tololo Inter-American Observatory/NSF NOIRLab, Casilla 603, La Serena, Chile}

\author{\OrcidIDName{0000-0001-5969-4631}{Y.~Zhang}}
\affiliation{NSF National Optical-Infrared Astronomy Research Laboratory, 950 N Cherry Avenue, Tucson, AZ 85719}

\author{\OrcidIDName{0000-0001-9789-9646}{J.~Zuntz}}
\affiliation{Institute for Astronomy, University of Edinburgh, Edinburgh EH9 3HJ, UK}

\date{\today}
\begin{abstract}
We present constraints on models of cosmology and astrophysics using cosmic shear data vectors from three datasets: the northern and southern Galactic cap of the Dark Energy Camera All Data Everywhere (DECADE) project, and the Dark Energy Survey (DES)  Year 3. These data vectors combined consist of 270 million galaxies spread across $13,\!000 \deg^2$ of the sky. We first extract constraints for \LCDM cosmology and find $\Seight = 0.805^{+0.019}_{-0.019}$ and $\Om = 0.262^{+0.023}_{-0.036}$, which is consistent within $1.9 \sigma$ of constraints from the \textit{Planck} satellite. Extending our analysis to dynamical dark energy models shows that lensing provides some (but still minor) improvements to existing constraints from supernovae and baryon acoustic oscillations. Finally, we study six different models for the impact of baryons on the matter power spectrum. We show the different models provide consistent constraints on baryon suppression, and associated cosmology, once the astrophysical priors are sufficiently wide. Current scale-cut approaches for mitigating baryon contamination result in a residual bias of $\approx 0.3\sigma$ in the $\Seight, \Om$ posterior. Using all scales with dedicated baryon modeling leads to negligible improvement as the new information is used solely to self-calibrate the baryon model on small scales. Additional non-lensing datasets, and/or calibrations of the baryon model, will be required to access the full statistical power of the lensing measurements. The combined dataset in this work represents the largest lensing dataset to date (most galaxies, largest area) and provides an apt testing ground for analyses of upcoming datasets from Stage IV surveys. The DECADE shear catalogs, data vectors, likelihoods, etc. are made publicly available.
\end{abstract}

\preprint{FERMILAB-PUB-25-0644-CSAID-PPD}
\maketitle

\section{Introduction}

\noindent Weak lensing (WL) is the deflection of light as it travels from distant sources to an observer. These deflections are sourced by the gravitational potential associated with the matter distribution between the sources and the observer \citep[see][for reviews]{Bartelmann2001, Schneider:2005:Lensing}. Given the direct connection to the matter distribution and to the distances between the observer, the matter distribution, and the sources, WL is a powerful probe of cosmological processes that affect both the growth of structure and the geometry of the Universe. Since its first detection nearly two decades ago \citep{Bacon2000, Wittman2000, Kaiser2000}, WL has been used extensively to constrain the cosmological parameters of our Universe \citep{Asgari2021, Amon2021, Secco2021, Li2023, paper4, Wright:2025:KiDsLegacy}. However, these measurements are also sensitive to a wide variety of extended cosmological models, such as modified gravity \citep[\eg][]{Schmidt:2008:MG_WL, DESY3KP2022}, extended inflationary models \citep[\eg][]{Maturi:2011:fNL, Shirasaki:2012:fNL, Anbajagane2023Inflation}, cosmological collider physics \citep[\eg][]{Goldstein:2024:inflation, Primordial1, Primordial2, Anbajagane:2026:tSZPNGs}, massive neutrinos \citep[\eg][]{Tereno:2009:CFHT_Mnu, Golshan:2025:Mnu}, dark matter phenomenology \citep[\eg][]{Preston:2025:DM} and in general, any process that alters the statistics of the cosmic matter distribution.

Here we present the next in a series of papers from the Dark Energy Camera All Data Everywhere (\decade) cosmic shear project. Previous works describe the assembly, validation, and calibration of the shear catalog (\href{\#cite.paper1}{Anbajagane \& Chang et al. \citeyear{paper1}}, hereafter \citetalias{paper1}), ensemble redshift estimates \citep[][hereafter \citetalias{paper2}]{paper2}, and analysis methodology (\href{\#cite.paper3}{Anbajagane \& Chang et al. \citeyear{paper3}}, hereafter \citetalias{paper3}). We then extracted cosmology constraints from a $\Lambda$ Cold Dark Matter (\LCDM) and \wCDM model (\href{\#cite.paper4}{Anbajagane \& Chang et al. \citeyear{paper4}}, hereafter \citetalias{paper4}). These works all focused on 5,400 $\deg^2$ of data from the northern Galactic cap (NGC). In the current work, we build on these efforts by extending the analysis to 3,400 $\deg^2$ of data in the southern Galactic cap (SGC). All data (both NGC and SGC) are processed in a manner that closely (or exactly) reflects the choices in the DES Y3 cosmic shear analysis \citep{y3-shapecatalog, Myles:2021:DESY3, Secco2021, Amon2021}. See other papers in the series for extended details on the analysis pipelines.

Using the improved precision afforded by these new datasets, we study WL-based constraints on models of both cosmology and astrophysics. Specifically, we study cosmology constraints from the \LCDM and dynamical dark energy models. We then constrain an astrophysics-based model for the suppression of the matter power spectrum on nonlinear scales, and study its coupling with cosmology in detail. The combination of \decade and DES Y3 data constitutes the largest weak lensing dataset to date --- $270$ million galaxies across $13,\!000 \deg^2$ of the sky. This combined dataset spans slightly more sky area, and has half the source galaxy number density, compared to the anticipated properties of the upcoming Year 1 dataset from the Vera C. Rubin Observatory Legacy Survey of Space and Time \citep[LSST, ][]{LSST2018SRD}. Thus, it offers a good testing ground for analyses that will be employed on the latter dataset.

This paper is organized as follows: we describe our dataset and analysis methods in Section~\ref{sec:data_methods}, including any relevant changes to the SGC pipeline relative to the NGC one presented in existing papers in this series. The extended models, and their corresponding parameter constraints, are shown in Section~\ref{sec:results}. We conclude in Section \ref{sec:summary}. We provide additional characterization of the SGC dataset in Appendix \ref{appx:SGC}, detail the baryon modeling pipeline in Appendix \ref{appx:BaryonModel}, and present our constraints on intrinsic alignments in Appendix \ref{appx:IA}.

\section{Dataset, Modeling and inference} \label{sec:data_methods}

We now briefly summarize the \decade dataset, our modeling choices, and our approach to parameter inference. Additional technical details can be found in the other papers in this series (\citetalias{paper1}, \citetalias{paper2}, \citetalias{paper3}, \citetalias{paper4}) and in the references therein.

\begin{figure*}
    \includegraphics[width = 2\columnwidth]{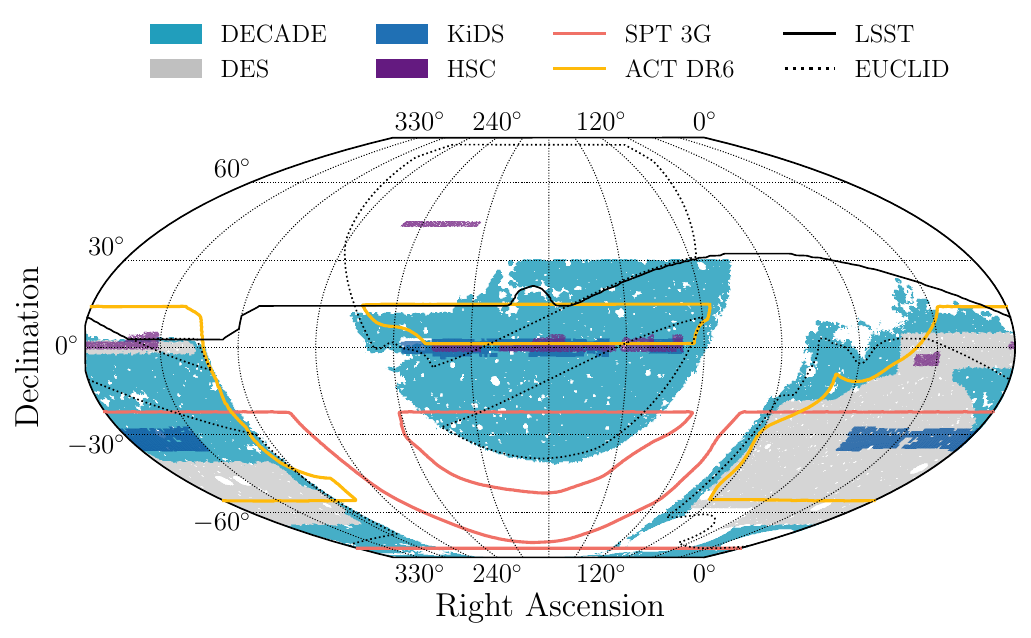}
    \caption{The footprint of the \decade cosmic shear analysis (light blue), in relation to those from other surveys. Reproduced from Figure 1 of \citetalias{paper4}, but now presenting the \decade SGC region as well. We show three other Stage-III surveys: DES Y3 (grey), KiDS-1000 (dark blue), and HSC Y3 (purple), and the footprints for the LSST wide-field survey (black solid), the \textit{Euclid} wide-field survey (black dotted), the SPT Ext-10k survey (orange), and ACT DR6 (yellow). See Section \ref{sec:data_methods} for references to the different experiments.} 
    \label{fig:footprint}
\end{figure*}

\begin{table}
    \centering
    \begin{tabular}{l c}
        Parameter & Prior \\
        \hline
        $\Omega_{\rm m}$ &  $\mathcal{U}(0.1,0.9)$ \\
        $\Omega_{b}$ &  $\mathcal{U}(0.03,0.07)$\\
        $h$ &  $\mathcal{U}(0.55,0.91)$ \\
        $A_s \times10^9$ &  $\mathcal{U}(0.5,5)$ \\
        $n_s$ & $\mathcal{U}(0.87,1.07)$ \\
        $\Omega_{\nu} h^2$ &  $\mathcal{U}(0.0006, 0.00644)$  \\
        \hline
        $a_{1}, a_{2}, \eta_{1}, \eta_{2}$ &  $\mathcal{U}(-4,4)$ \\
        $b_{\rm ta}$  & $\mathcal{U}(0,2)$ \\       
        \hline
        $\Delta z_1 \times 100$  &$\mathcal{N}(0, 1.63)$ \& $\mathcal{N}(0, 1.61)$ \\
        $\Delta z_2 \times 100$  &$\mathcal{N}(0, 1.39)$ \& $\mathcal{N}(0, 1.40)$ \\
        $\Delta z_3 \times 100$  &$\mathcal{N}(0, 1.01)$ \& $\mathcal{N}(0, 1.00)$ \\
        $\Delta z_4 \times 100$  &$\mathcal{N}(0, 1.17)$ \& $\mathcal{N}(0, 1.16)$ \\
        \hline
        $m_1 \times 100$ & $\mathcal{N}(-0.92, 0.296)$ \& $\mathcal{N}(-1.33, 0.472)$  \\
        $m_2 \times 100$ & $\mathcal{N}(-1.90, 0.421)$ \& $\mathcal{N}(-2.26, 0.657)$ \\
        $m_3 \times 100$ & $\mathcal{N}(-4.00, 0.428)$ \& $\mathcal{N}(-3.67, 0.697)$ \\
        $m_4 \times 100$ & $\mathcal{N}(-3.73, 0.462)$ \& $\mathcal{N}(-5.72, 0.804)$ \\
        \hline
    \end{tabular}
    \caption{Cosmological and nuisance parameters in the baseline \LCDM model. Uniform distributions in the range $[a,b]$ are denoted $\mathcal{U}(a,b)$ and Gaussian distributions with mean $\mu$ and standard deviation $\sigma$ are denoted as $\mathcal{N}(\mu,\sigma)$. We show both NGC \& SGC priors for the redshift and shear nuisance parameters. The IA parameters are independent for each of the three data vectors (\decade NGC, \decade SGC, and DES Y3). The priors on the calibration nuisance parameters for DES Y3 can be found in \citet{Secco2021} and \citet{Amon2021}}
    \label{tab:params}
\end{table}

\subsection{Data}\label{sec:data_model:data}

Our main measurements are the angular two-point correlation functions of galaxy orientations, denoted as $\xi_\pm$. See \citetalias{paper4} for details on the estimator for this statistic. Our analysis uses three separate cosmic shear data vectors --- \decade NGC, \decade SGC, and DES Y3. The \decade NGC and DES Y3 data vectors were already combined in the analysis of \citetalias{paper4}. In this work we further supplement this by adding measurements from the southern Galactic cap (SGC). This adds another $3,\!356 \deg^2$ of sky surrounding DES, with 63 million galaxies in total. We define the footprint and sample selections using the same procedures as in \citetalias{paper1}; the sole new addition is a $20\arcdeg$ and $10\arcdeg$ aperture mask around the Large and Small Magellanic Clouds, respectively. The SGC region is defined to be independent of the DES Y3 footprint. See Figure \ref{fig:footprint} for the final area covered by the combined \decade data. Figure \ref{fig:datavectors} in Appendix \ref{appx:DV} shows the data vectors from all three datasets.

All data products from the \decade cosmic shear project, including catalogs, data vectors, and likelihoods, are now publicly accessible. The broader set of photometric data --- denoted as Data Release 3 (DR3) from the Dark Energy Camera Local Volume Exploration survey \citep[DELVE,][]{Drlica-Wagner:2021:DELVE} --- will be presented in \textcolor{blue}{Drlica-Wagner et al. (in prep)}. Details on the associated image processing can also be found in \citet{Tan:2025:LeoVI}.

We have processed the additional SGC region through all calibration pipelines and tests presented in the \decade cosmic shear project \citepalias{paper1, paper2, paper3}. We have also done all pre-unblinding tests as discussed in Appendix C of \citetalias{paper4} and confirmed all tests pass for the SGC region. The constraints from the SGC are presented in Appendix \ref{appx:SGC}, and are consistent with those of the NGC (and DES). Therefore, we can derive constraints from combining all three data vectors. This combination is done at the likelihood level --- that is, without modeling any cross-covariance between surveys --- since the NGC, SGC, and DES Y3 datasets cover independent patches of the sky. We have also redone the shear and redshift calibrations using the same methods as \citetalias{paper1} and \citetalias{paper2}, respectively, and list the associated nuisance parameters below in Table \ref{tab:params}. The redshift distributions are listed in Figure \ref{fig:RedshiftsSGC} of Appendix \ref{appx:SGC}.

In this work, we also consider two additional datasets:
\begin{itemize}
    \item Measurements of the Baryon Acoustic Oscillations (BAO) from Data Release 2 of the Dark Energy Spectroscopic Instrument \citep[][see their Table IV]{DESI:2025:BAO}. In total, there are 13 measurements of the BAO scaling parameter and distance ratios. We will henceforth refer to this data as DESI DR2.\vspace{8pt}
    
    \item The DES Y5 Supernovae (SNe) sample \citep{DES:2024:SNe}, supplemented by a historical sample of low-redshift SNe from Cfa3 \citep{Hicken:2009:SNe}, Cfa4 \citep{Hicken:2012:SNe}, the Carnegie SNe Project \citep{Krisciunas:2017:SNe}, and the Foundation SNe Survey \citep{Foley:2018:SNe}. The combined sample contains 1829 SNe between $0.01 < z < 1.1$, with 1635 SNe from the DES sample spanning $0.1 < z < 1.1$. The effective number of SNe magnitude measurements is 1735 as mentioned in Section 4.2.1 of \citet{DES:2024:SNe}.\vspace{8pt}
\end{itemize}
The published analyses of the above two probes use slightly different priors than the ones used in this work \citep{DES:2024:SNe, DESI:2025:BAO}. However, as we discuss in Section \ref{sec:darkenergy}, our cosmology constraints from the combination of BAO and SNe are consistent (within sampling noise) with the published results. Hence this difference in priors is inconsequential in practice.

We do not consider Cosmic Microwave Background (CMB) measurements in this analysis, in order to retain the low-redshift nature of our combined constraints. In our analysis, we obtain geometry (distance) information from DESI BAO and from DES SNe, and obtain growth (density perturbation) information from WL. The latter probe also contributes geometric information as its signal is sensitive to cosmological distances between the source galaxy, the observer, and the intervening matter, but for this particular combination of probes its biggest contribution is its information on the growth of structure.

One noteworthy detail is that once we combine the DECADE NGC and SGC regions with DES Y3, the resulting catalog effectively triples the sky coverage of precision weak lensing datasets. This provides significantly more overlap between such weak lensing datasets and other wide-field cosmological surveys, including CMB experiments like \textit{Planck} \citep{Planck:2020:LegacyOverview}, the South Pole Telescope \citep[SPT,][]{Carlstrom2011}, the Atacama Cosmology Telescope \citep[ACT,][]{ACT:2007, ACT:2016}, and the Simons Observatory \citep[SO,][]{Simons:2019:Experiment}; spectroscopic datasets such as the Sloan Digital Sky Survey \citep{York_2000, Dawson_2013, Dawson_2016} and DESI; as well as X-ray surveys like eROSITA \citep{Merloni:2012:Erosita}. While this work is limited to a cosmic shear analysis, the dataset presented here is conducive to many cross-correlation analyses that can further stress-test the cosmology constraints and probe a variety of astrophysical and cosmological questions \citep[\eg][]{Shin:2019:Splashback, Gatti2021DESxACT, Pandey2021DESxACT, Tilman:2022:tSZxWL, Chang2023, Omori:2023:CMBL, Sanchez:2023:tSZ, Anbajagane:2024:Shocks, Bigwood:2024:BaryonsWLkSZ}.

\subsection{Modeling and inference}\label{sec:data_model:model}

Our modeling choices are described in detail in \citetalias{paper3} and \citetalias{paper4}. In brief, our pipeline follows that of DES Y3 \citep{Krause2021}. The only difference is we now use \textsc{HMCode} \citep{Mead2020a} as our model for the nonlinear matter power spectrum, following the choice adopted in the DES \& Kilo Degree Survey (KiDS) joint reanalysis \citep{DESKiDS2023}.

Since the weak lensing (or shear) effect is sourced by the matter distribution, the shear two-point correlations can be predicted from the matter two-point correlations, \ie from the matter power spectrum. In specific, the shear correlations can be modeled as,
\begin{align} \label{eq:xipm}
    \xi_{\pm}^{ij}(\theta)= \sum_{\ell} & \,\,\frac{2\ell+1}{2\pi\ell^{2}\left(\ell+1\right)^{2}}\left[G_{\ell,2}^{+}\left(\cos\theta\right)\pm G_{\ell,2}^{-}\left(\cos\theta\right)\right]\nonumber\\& \times\left[C_{EE}^{ij}(\ell)\pm C_{BB}^{ij}(\ell)\right],
\end{align}
where the functions $G^\pm_\ell(x)$ are computed from Legendre polynomials $P_\ell(x)$ and averaged over angular bins \citep{Krause2021}. The $i$ and $j$ indices specify the two tomographic redshift bins from which the correlation function is calculated. The term $C_{EE}$ contains the angular matter power spectrum integrated along the line-of-sight after being weighted by the lensing kernels (see Equation 2 in \citetalias{paper3}).

The intrinsic alignments (IA) of galaxies also contribute to the $C_{EE}$ and $C_{BB}$ terms. This IA signal is connected to the matter distribution and so spatial correlations of galaxy IA can be predicted using the matter power spectrum as well. We model IA using the Tidally Aligned Tidally Torqued \citep[TATT,][]{Blazek2019} approach, following \citet{Secco2021} and \citet{Amon2021}. The amplitude of this IA contribution is parameterized as,
\begin{align}\label{eqn:IA_ampl}
    A_1(z) &= -a_1 \bar{C}_1 \frac{\rho_{\text{crit}} \Om}{D(z)} 
    \left( \frac{1 + z}{1 + z_0} \right)^{\eta_1}, \\
    A_2(z) &= 5a_2 \bar{C}_1 \frac{\rho_{\text{crit}} \Om}{D^2(z)} 
    \left( \frac{1 + z}{1 + z_0} \right)^{\eta_2},\\
    A_{1\delta}(z) &= b_{\rm TA} A_1(z),
\end{align}
where $A_1$ and $A_2$ scale the matter power spectra, $D(z)$ is the linear growth rate, $\rho_{\text{crit}}$ is the critical density at $z = 0$, and $\bar{C}_1 = 5 \times 10^{-14} M_\odot h^{-2} \text{Mpc}^2$ is a normalization constant, set by convention. We choose a pivot redshift, $z_0 = 0.62$ following existing work \citep[][\citetalias{paper4}]{Secco2021, Amon2021}. The free parameters of our model are the amplitudes $a_1, a_2, b_{\rm TA}$ and the power-law indices $\eta_1, \eta_2$. See Equations 20--23 in \citet{Secco2021} for a description of the different IA-related power-spectra that contribute to the final signal.

We fit the model above to our $\xi_{\pm}$ measurements using a Markov Chain Monte Carlo (MCMC) approach. We assume a Gaussian likelihood $L$, with 
\begin{equation} \label{eq:likelihood} 
    \ln L ( \xi_{\pm,d} | \boldsymbol{p})
    = -\frac{1}{2}\bigg(\xi_{\pm,d} - \xi_{\pm,m}(\mathbf{p})\bigg)\mathbf{C}^{-1} \bigg(\xi_{\pm,d} - \xi_{\pm,m}(\mathbf{p})\bigg), 
\end{equation} 
where $\xi_{\pm}$ is a concatenation of the $\xi_+$ and $\xi_-$ measurements; $\xi_{\pm,d}$ and $\xi_{\pm,m}$ are the data vectors measured in the data and predicted from our theoretical model; $\textbf{C}^{-1}$ is the inverse covariance of the measurements; $\textbf{p}$ is a vector of the cosmology parameters and nuisance parameters listed in Table~\ref{tab:params}. The Bayesian posterior is proportional to the product of the likelihood $L$ and the prior $P$, or 
\begin{equation} \label{eqn:evidence}
    P(\mathbf{p}|\xi_{\pm,d}, M) =  \frac{\mathcal{L} ( \xi_{\pm,d} | \mathbf{p}, M)P(\mathbf{p} | M)}{P( \xi_{\pm,d} | M)},
\end{equation} 
where the denominator, $P( \xi_{\pm,d} | M)$, is the ``evidence'' of the data. The entire expression is conditioned on a model choice, $M$. We note that the evidence is sensitive to choices of the prior, so we only use it as a general guideline for model preference, and not as any precise, quantitative measure.

Our parameter prior choices are listed in Table~\ref{tab:params}. The nuisance parameters for DES Y3 are taken from Table 1 of \citet{Secco2021}. We generate a covariance matrix for the data vectors using \textsc{CosmoCov} \citep{Krause:2017:CosmoLike, Fang2020, Fang2020b}, and our approach follows the methods of \citet{Friedrich:2021:CovY3} as employed in DES Y3. The final covariance matrix comprises of a simple Gaussian covariance term, as well as a connected four-point term to account for nonlinear structure \citep{Wagner:2015:Response, Barreira:2017:Response, Barreira:2017:ResponseCov}, a super-sample contribution to incorporate correlations between small-scale modes as generated by modes larger than the survey footprint \citep[\eg][]{Barreira:2018:SSC}, and also a correction for the impact of the survey mask on the shape noise term \citep{Troxel:2018:Cov}. Following \citetalias{paper4}, we compute the covariance (and perform inference) in an iterative fashion: we first compute the covariance at the best-fit cosmology from \citetalias{paper4}. We perform inference on the combination of DECADE NGC, DECADE SGC, and DES Y3 (the ``DECam 13k'' result in Table \ref{tab:constraints}) and recompute recompute the covariance at this new best-fit cosmology. All presented results are from inference performed with this updated covariance.

All parameter inference is performed using the \textsc{CosmoSIS} package \citep{Zuntz2015}. One key change in this work, relative to \citetalias{paper4}, is the use of \textsc{Polychord} \citep{Handley:2015:Polychord} as our fiducial sampler. While we used the \textsc{Nautilus} sampler \citep{Lange:2023:Nautilus} in \citetalias{paper4}, we find its runtime is significantly longer for our combined analysis of three data vectors with its increased parameter dimensionality. The steep scaling of the \textsc{Nautilus} runtime with parameter dimensionality is also discussed in \citet[][see the end of their Section 4.3]{Lange:2023:Nautilus}. After initial attempts with \textsc{Nautilus}, we chose to use the \textsc{Polychord} sampler, which we have verified produces the same posteriors as \textsc{Nautilus} \citepalias{paper3, paper4}. The hyper-parameters choices for the samplers are listed in Table 2 of \citetalias{paper3}.

Finally, when performing a joint analysis across the NGC, SGC, and DES Y3 data vectors, there are two possible approaches for the IA modeling: first, we can use an independent set of IA parameters for each survey, resulting in 15 free parameters under the TATT model. Second, we can use a common set of parameters for both surveys, resulting in 5 free parameters under the TATT model. We use the former, more conservative choice for our analysis. The two approaches were explored for the combination of DECADE NGC and DES Y3 as presented in \citetalias{paper4}, and were found to give consistent constraints on cosmology.

\section{Results} \label{sec:results}

\begin{figure*}
    \centering
    \includegraphics[width=\columnwidth]{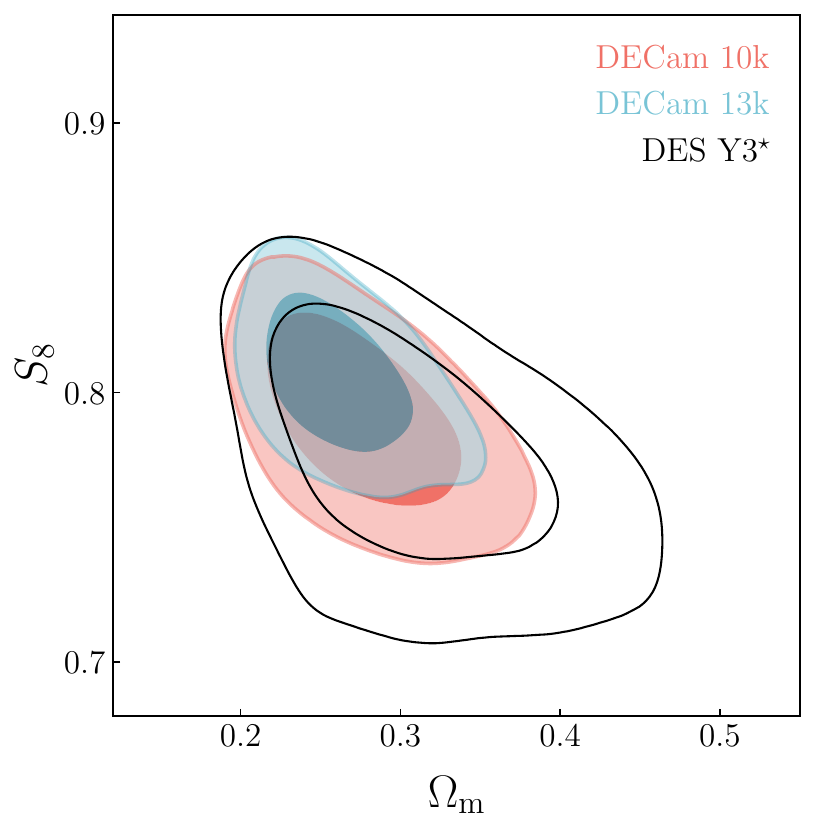}
    \includegraphics[width=0.987\columnwidth]{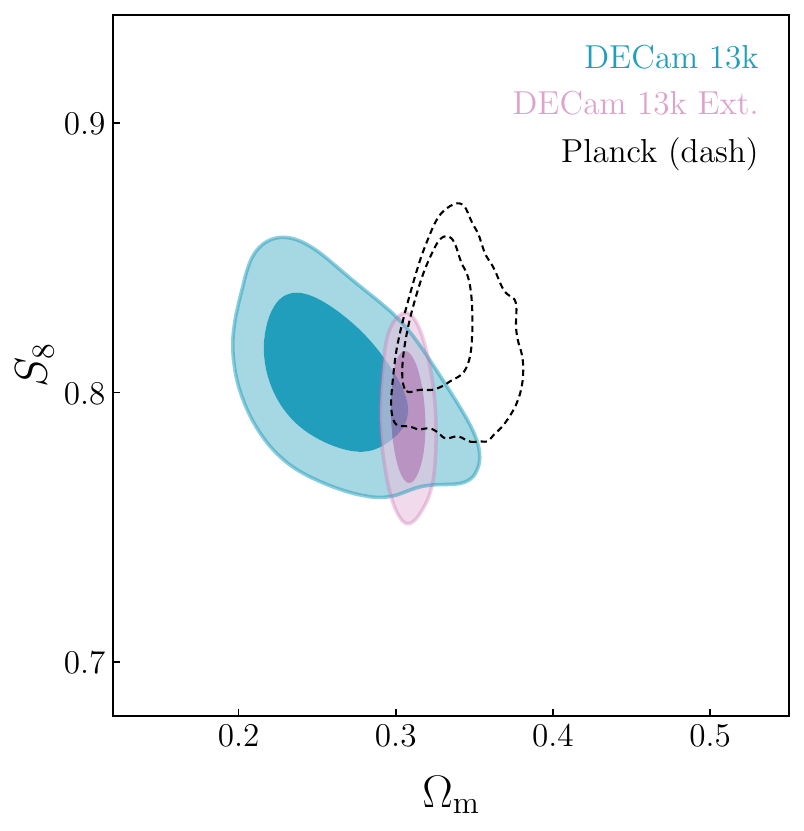}
    \caption{The \LCDM constraints from \decade and DES Y3. We refer to the combination of \decade NGC, \decade SGC, and DES Y3 as ``DECam 13k'', and that of just \decade NGC and DES Y3 as ``DECam 10k''. We show the latter for easier comparisons with the results of \citetalias{paper4} and to highlight the improvement from the inclusion of the \decade SGC data. The full 13,000 $\deg^2$ (13k) dataset has a Figure of Merit that improves by three times over the DES Y3-only result (Table \ref{tab:constraints}). The combination with external data (DESI DR2 BAO and DES Y5 SNe) constrains $\Om$ and prefers lower values of $\Seight$, which are still consistent with the posterior of the 13k shear-only analysis.}
    \label{fig:LCDM}
\end{figure*}

All numerical constraints quoted below are the mean of the parameter posterior with the 16\% and 84\% percentiles, and all plotted contours show the 1$\sigma$ and 2$\sigma$ regions. Following other work in the lensing community, we quote constraints on the derived parameter $\Seight \equiv \sigma_8 (\Om/0.3)^{0.5}$ and also list the ``Figure of Merit'' (FoM) of our posteriors, computed in the $\Seight- \Om$ plane as
$\text{FoM}_{\Seight\Om} = \det[\text{Cov}(\Seight, \Om)]^{-1/2}$.

One aspect of our main results is to quantify the agreement/disagreement between different constraints. There are a variety of methods for estimating the significance of a shift between two contours \citep[\eg][]{Lemos:2021:Tensions}. We use the Gaussian tension metric within the \textsc{tensiometer}\footnote{\url{https://github.com/mraveri/tensiometer}} code \citep{Raveri:2021:NGTension}. In all \LCDM analyses to follow, we compute the tension in the $\Seight$-$\Om$ plane of the two datasets, and we find the posterior in this plane can be closely approximated by a multivariate Gaussian distribution. When computing significances for the dynamical dark energy models, we evaluate the tension between a posterior and a single point (the \LCDM value, \eg $w = -1$).

Throughout, we refer to the combination of \decade NGC, \decade SGC, and DES Y3 as ``DECam 13k'', and that of \decade NGC and DES Y3, which was carried out in \citetalias{paper4}, as ``DECam 10k''. The latter is presented here to better compare with \citetalias{paper4} and highlight the contributions from the DECADE SGC data. All analyses with DES Y3 only consider the cosmic shear data vector and ignore the shear ratio measurements, following \citet{DESKiDS2023} and \citetalias{paper4}. Results from combinations with BAO and SNe measurements will be denoted with ``Ext.''

As mentioned before, our analysis pipeline follows the choices listed in \citetalias{paper4}. We have verified that prior volume effects cause minimal/subdominant shifts in the marginal posteriors of the parameters of interest. Constraints from the independent \decade SGC region alone are detailed in Appendix \ref{appx:SGC}, and those on the intrinsic alignments parameters are in Appendix \ref{appx:IA}. The summary of dynamical dark energy results are shown in Table \ref{tab:dynamDE}. The same for all \LCDM results, including the addition of baryon modeling, is shown in Figure \ref{fig:Summary} and tabulated in Table \ref{tab:constraints}.

\subsection{Base analysis: $\bold{\Lambda}$CDM}\label{sec:results:LCDM}

We start by analysing the data under the standard \LCDM paradigm. We obtain constraints of,
\begin{align} \label{eqn:LCDM}
    S_8 = & \,\, 0.805\pm 0.019, \notag\\
    \Omega_{\rm m} = & \,\,0.262^{+0.023}_{-0.036},
\end{align}
with the 2D posteriors shown in Figure \ref{fig:LCDM}.

Relative to our results in \citetalias{paper4}, the inclusion of additional weak-lensing data pushes the contour further towards lower values of $\Om$. This is already seen by the comparison in \citetalias{paper4} (see their Figure 2) of the \decade NGC constraints with those of \decade NGC combined with DES Y3. The resulting contour is still consistent with \textit{Planck} 2018 \citep{Planck:2020:Cosmo}. While there are slight differences in the $\Om$ dimension, the contours are consistent within $1.9\sigma$ in the $\Om - \Seight$ plane. Adding the BAO and SNe data alongside WL improves the consistency with \textit{Planck}  2018 in $\Om$, as the former probes cause the posterior to no longer prefer any low values of $\Om$. The combined constraint is consistent with \textit{Planck} 2018 at $1.5\sigma$. Note that the presented \textit{Planck} 2018 results are from our reanalysis of the temperature and polarization measurements (denoted as the ``TT,TE,EE+lowE'' likelihood in \citet{Planck:2020:Cosmo}), but now using the same cosmology priors employed for the weak lensing analysis. This follows our approach in \citetalias{paper4}. Note that the lensing analyses from DES and from this work vary the neutrino mass within a finite prior (Table \ref{tab:params}) whereas the fiducial \textit{Planck} results fix it to $\Omega_\nu h^2 = 0.000644$ (approximately the lower bound of our prior). Applying the lensing priors to the \textit{Planck} analysis results in the latter's constraints extending to larger values of $\Om$, and therefore differing from the fiducial \textit{Planck} $\Lambda$CDM constraints. As a result, our \textit{Planck} contour in Figure \ref{fig:LCDM} differs from the public $\Lambda$CDM chain.

Table \ref{tab:constraints} lists the constraints of the different datasets, alongside their associated best-fit $p$-values. The $p$-value of the DECam 13k \LCDM best-fit is $p = 0.007$. While this formally passes the $p > 0.0015$ ($3\sigma$) threshold set in \citetalias{paper4}, it is still low. In Section 5.1 of \citetalias{paper4}, we found the \decade NGC dataset has three tomographic bin combinations where the $\xi_+$ measurements have a larger $\chisq$ (resulting in $p = 0.017$ for the full data vector). Dropping these bins causes no change to any of the cosmology or nuisance parameter posteriors, but improves the $p$-value significantly (to $p = 0.56$ for the remaining data points). We also confirmed in \citetalias{paper4} that the large $\chisq$ was not correlated with any survey property maps. 

In this work, we rerun our DECam 13k analysis after dropping these three bin combinations in \decade NGC. We confirm that the parameter posteriors show no visible change but the $p$-value now increases to $p = 0.14$. The constraints on cosmology --- $\Seight =  0.805\pm 0.021$ and $\Om = 0.259^{+0.022}_{-0.042}$ --- are fully consistent with the fiducial constraints, and the two posteriors are within $<0.01\sigma$ in the $\Seight-\Om$ plane. In summary, the DECam 13k analysis exhibits a low $p$-value that still passes our pre-determined criteria. The origin of this lower value is in just the \decade NGC dataset, and we confirm that dropping the three tomographic bin combinations with the largest $\chisq$ has a completely negligible impact on the final cosmology constraints (including for the extended models we study in this work). The resulting cosmology constraints in all cases are consistent with \textit{Planck} 2018. We also note that all other variant analyses of DECam 13k exhibit much higher values of $p > 0.01$, and most are at $p \approx 0.1$ (Table \ref{tab:constraints}).

Table \ref{tab:constraints} also quotes the FoM for all the different \LCDM analyses. The DECam 13k dataset improves on the DES Y3 constraints by more than a factor of 3 in the FoM. As noted earlier, the combined dataset spans slightly more sky area ($\approx$ 13,000 $\deg^2$), and has half the source galaxy number density, compared to the anticipated properties of the upcoming Year 1 LSST dataset \citep{LSST2018SRD}.

The IA constraints for the joint analysis are consistent with those obtained from analyzing the individual data vectors on their own. This is expected since we have modeled the IA contribution in each data vector using a separate set of IA parameters (see Section \ref{sec:data_model:model}). The constraints from DECADE NGC are discussed in detail in Section 5.2 of \citetalias{paper4} and those from DES Y3 in \citet{Secco2021}. The SGC IA results, which are new to this work, are statistically consistent with no IA, \ie $a_1 = 0$ and $a_2 = 0$. Further discussion on all IA results can be found in Appendix \ref{appx:IA} and Figure \ref{fig:IA}.

\subsection{Dynamical dark energy}\label{sec:darkenergy}

\begin{figure*}
    \centering
    \includegraphics[width=\columnwidth]{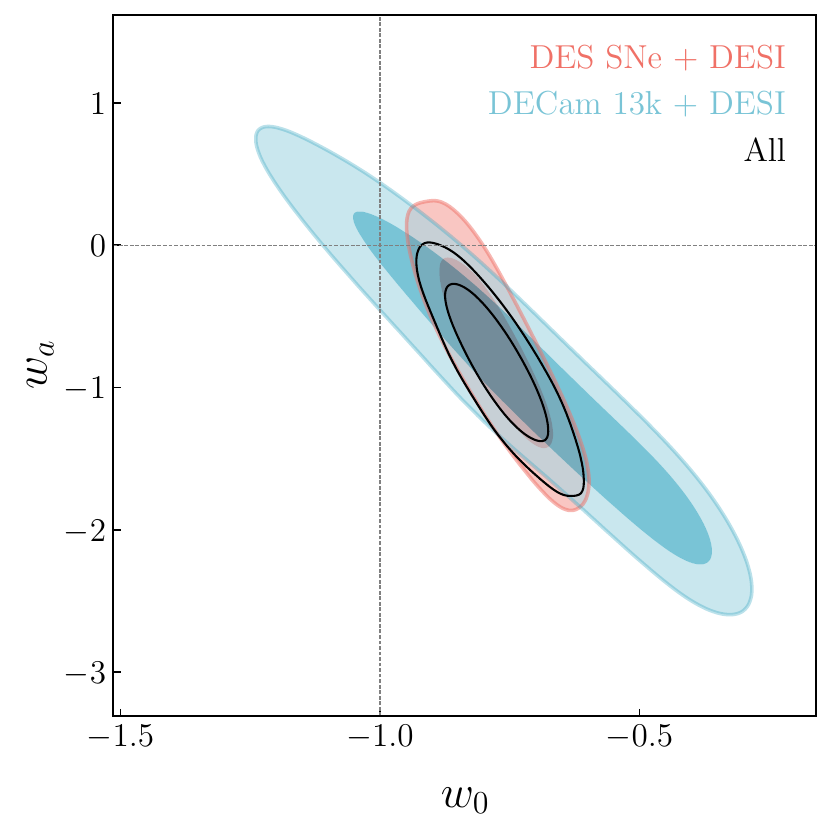}
    \includegraphics[width=\columnwidth]{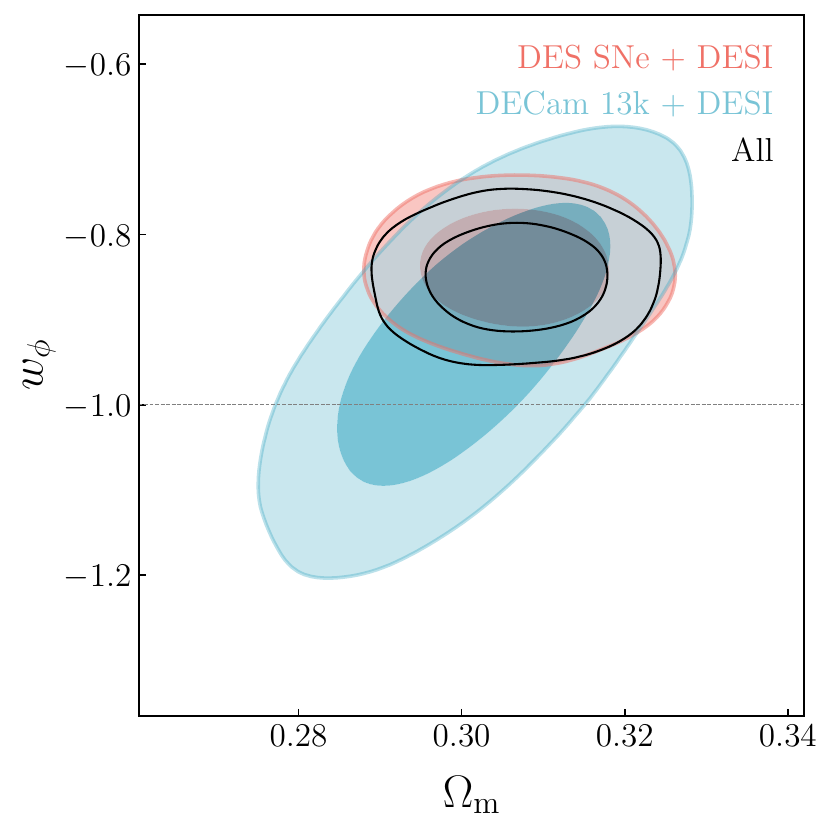}
    \caption{The constraints on two dynamical dark energy models --- the phenomenological $w_0w_a$ model (left) and a first-principles scalar field model from \citet{Shajib:2025:wphi} (right). In both cases, the combination of cosmic shear plus BAO does not find any deviation from \LCDM, while combining BAO and SNe does show deviations. Adding lensing to this latter combination marginally improves the constraints. See Table \ref{tab:dynamDE} for numerical values.}
    \label{fig:darkenergy}
\end{figure*}

{\setlength{\tabcolsep}{5.5pt}
\begin{table*}
    \centering
    \begin{tabular}{c|c|cccc|ccc}
    \hline
    Model & Run & $w_0 (w_\phi)$ & $w_a$ & $\Om$ & $\alpha$ & $\sigma$(\LCDM) & $\chi^2/N_{\rm dof}$ & $p$-value \\
    \hline
    \rule{0pt}{10pt}
    & BAO + SN & $-0.777\pm 0.071$ & $-0.75\pm 0.45$ & $0.320^{+0.016}_{-0.012}$ &--- &  $3.2$ & $1655/1748$ & 0.944  \\[5pt]
    $w_0w_a$& WL + BAO & $-0.71^{+0.17}_{-0.28}$ & $-0.99^{+0.91}_{-0.70}$ & $0.327^{+0.031}_{-0.022}$& ---  & $0.8$ & $780/694$ & 0.013  \\[5pt]
    & WL + BAO + SN & $-0.773^{+0.060}_{-0.069}$ & $-0.84^{+0.39}_{-0.34}$ & $0.322\pm 0.011$& ---  & $3.3$ &  $2428/2429$ & 0.502 \\[5pt]
    \hline 
    \rule{0pt}{10pt}
    & BAO + SN & $-0.839\pm 0.045$ & --- & $0.3068\pm 0.0077$ & $1.451\pm 0.058$  & $3.6$ & $1656/1748$ & 0.942 \\[5pt]
    $w_\phi$& WL + BAO & $-0.93^{+0.11}_{-0.10}$ & --- & $0.301\pm 0.011$ & $1.447\pm 0.058$ &  $0.66$ & $782/694$ & 0.010  \\[5pt]
    & WL + BAO + SN & $-0.851\pm 0.042$ & --- & $0.3067\pm 0.0073$ & $1.454^{+0.085}_{-0.042}$ &  $3.5$ &  $2430/2429$ & 0.490 \\[5pt]
    \hline
    \end{tabular}
    \caption{Constraints on two dynamical dark energy models, $w_0w_a$ and $w_\phi$, for three different combinations of datasets as shown in Figure \ref{fig:darkenergy}. The two models use a redshift-evolving equation-of-state, $w(z) = w_0 + zw_a/(1 + z)$ and $w(z) = -1 + (1 + w_\phi)e^{-\alpha z}$, respectively. The constraints are shown in Figure \ref{fig:darkenergy}. The $w_\phi$ model has a free parameter $\alpha$ that is marginalized over a theory-informed, uniform prior of $1.35 < \alpha < 1.55$.
    }
    \label{tab:dynamDE}
\end{table*}
}

Cosmological constraints from DES SNe \citep{DES:2024:SNe} and DESI BAO \citep{DESI:2025:DR1BAO, DESI:2025:BAO}, when combined with other external data, hint at a potential redshift evolution in the energy density of dark energy. We consider two such parameterizations --- the $w_\phi$ model \citep{Shajib:2025:wphi} and $w_0w_a$ model \citep{Chevallier:2001:w0wa, Linder:2003:w0wa} --- and show constraints below. The former is a physically motivated class of trajectories for the evolution of this energy density over redshift. In particular, all models in this class strictly follow the null energy condition (NEC) for which the sum of the pressure and the energy density of dark energy is non-negative. 
The latter model is a generic, parametric approach that was constructed to probe any evolution in $w(z)$ without being tied to a physical model. The two models use a redshift-evolving equation-of-state, $w(z) = -1 + (1 + w_\phi)e^{-\alpha z}$ and $w(z) = w_0 + zw_a/(1 + z)$, respectively. We use priors of $w_\phi \in [-2, -1/3]$ and $\alpha \in [1.35, 1.45]$ for the former, and $w_0 \in [-2, -1.3]$ and $w_a = [-3, 3]$ for the latter. The prior on the parameter $\alpha$ is informed by theoretical considerations \citep[][see their Section 2]{Shajib:2025:wphi}.

\begin{figure*}
    \centering
    \includegraphics[width=2\columnwidth]{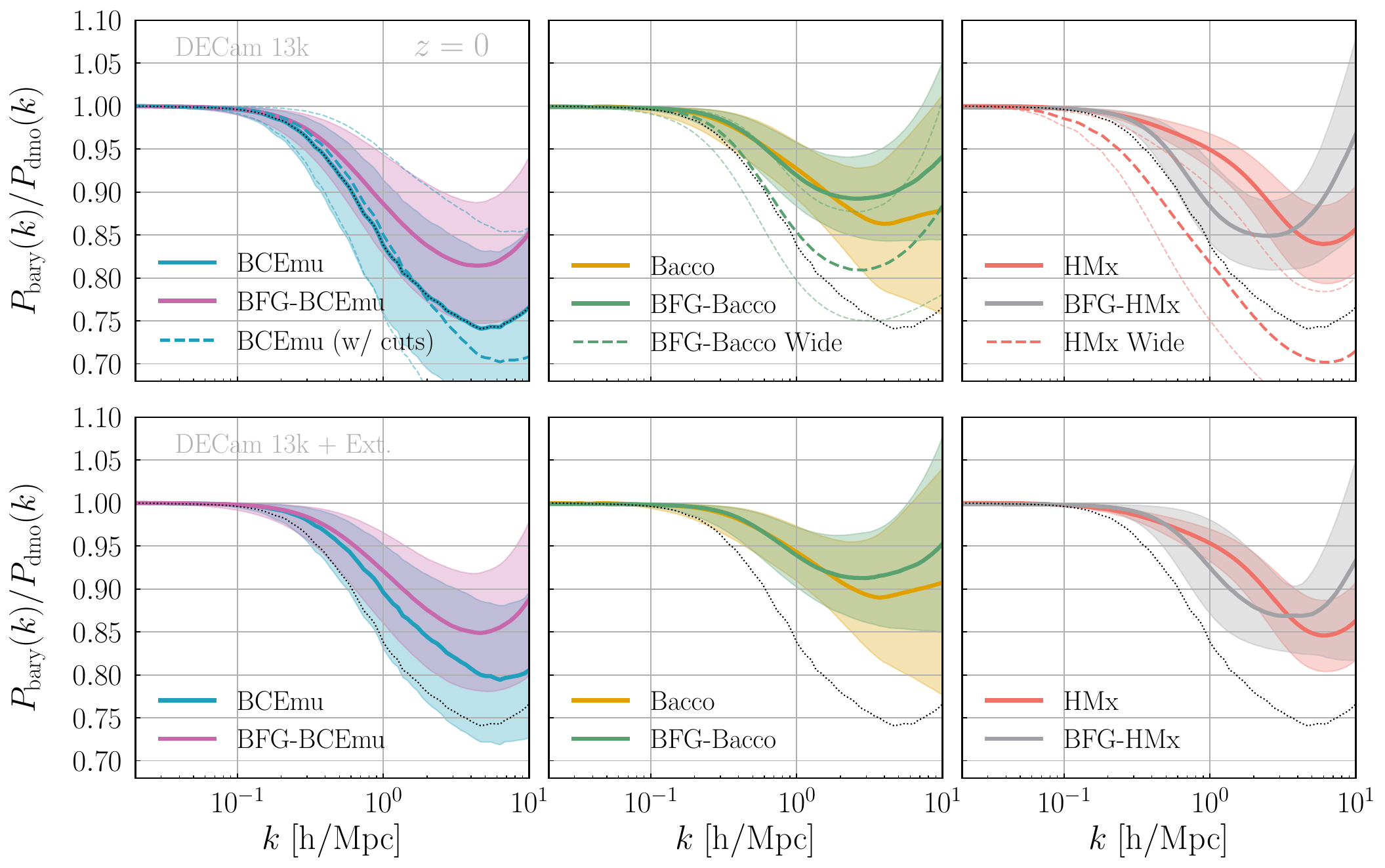}
    \caption{The baryonic suppression at $z = 0$ from the different models. We show three popular methods --- \BCEmu, \Bacco, and \HMx. For each, we provide a version obtained from a halo-model calculation using the \textsc{BaryonForge} pipeline. See text for details. The black dotted line in each panel is the \BCEmu result for the DECam 13k dataset, shown as reference. In general, the BCEmu-derived matter power suppression is stronger than those from the other models. However, concordance can be achieved by adequately widening the priors of the other models (the models tagged with ``Wide'' in the above plot). See Appendix \ref{appx:BaryonModel} for more details on the modeling and the choice of priors. The BFG-\BCEmu results matches the \BCEmu result if the former adopts the narrower cosmology priors of the latter (Figure \ref{fig:Testfixcosmo}). The inclusion of external data, from DESI DR2 BAO and DES Y5 SNe (bottom), provides results that are consistent with the fiducial case.}
    \label{fig:PkSuppression}
\end{figure*}

In both parameterizations, the majority of the constraining power comes from the BAO and SNe results, with weak lensing providing some, but still minor, improvements. We present these results in Figure \ref{fig:darkenergy}. The left panel shows the $w_0w_a$ parameterization, where lensing provides noticeable improvements to the constraints from the extended datasets (BAO and SNe) but in a direction mostly orthogonal to the best constrained direction of the contour. As a result, the significance of the deviation from \LCDM ($3.1\sigma$, for the combined analysis) increases only slightly, to $3.2\sigma$. A similar result is found for the $w_\phi$ model, where lensing does improve the constraining power but not significantly. In the $w_\phi$ case the improvement from adding WL is more marginal than in the $w_0w_a$ case as the $w_\phi$ class of models already spans a narrower range of predictions --- it incorporates a theoretical prior (namely, the dynamical dark energy field must not violate the NEC) that reduces the space of predictions --- and so the WL information provides less improvement. We summarize these results in Table \ref{tab:dynamDE}. Our results from the combination of just BAO and SNe is consistent with \citet[][see their Table V]{DESI:2025:BAO} for the $w_0w_a$ model, and \citet[][see their Table II]{Shajib:2025:wphi} for the $w_\phi$ model. Upon computing the ratio of the Bayesian evidences (Equation \ref{eqn:evidence}) from each extended model with respect to LCDM, we find ratios of $R_{w_0w_a} = 25.8$ and $R_{w_\phi} = 18.8$, which indicates a preference towards these models over LCDM. This is expected given the deviation ($3.2\sigma$) from LCDM as noted above. Our values for $R_{w_\phi}$ are largely consistent with those found in Table II of \citet{Shajib:2025:wphi}, and are dominated by information from SNe and BAO.

\citet{DESI:2025:BAO} also present results where combining with DES Y3 shows considerable improvement on the $w_0w_a$ constraints. That analysis used measurements of cosmic shear, as well as those of galaxy clustering and galaxy-galaxy lensing \citep{DESY3KP2022}. Compared to the former probe, the latter two probes have an enhanced sensitivity to the time evolution of the expansion rate, $H(z)$, and so provide better constraints on $w(z)$. For a similar reason, they also constrain $\Om$ more precisely, and this parameter has been linked to the observed deviations in the dark energy parameter space \citep{Tang:2025:OmDE}.

Note that the measured deviation from \LCDM can vary depending on the exact choice of SNe sample used in the analysis \citep[][see their Figure 11]{DESI:2025:BAO}. In particular, combining DES Y5 SNe with BAO is known to cause larger deviations from \LCDM relative to that obtained by combining BAO with other SNe samples like Pantheon$+$ \citep{Scolnic:2022:Pantheon, Brout:2022:Pantheon} or UNIONS \citep{Rubin:2025:Unions}. In our work, we are interested in quantifying the improvement to the dynamical dark energy constraints due to the addition of weak lensing from $13,\!000\deg^2$. Given this goal, we do not analyze all different permutations over SNe samples and utilize just DES Y5 SNe.

\subsection{Baryon signatures}\label{sec:Baryons}

We now use the three cosmic shear data vectors (without applying any scale cuts) and account for the impact of baryon imprints on nonlinear scales via various phenomenological models.\footnote{Our covariance model follows that of \citetalias{paper3}, and does not account for baryonic effects. Such effects will matter only on small scales, where the covariance is completely dominated by shape noise (the random orientations of intrinsic galaxy shapes) rather than cosmic variance. Hence, these effects can be ignored in the covariance model.} A wide array of such models have been used in the literature: \Hmx \citep{Mead2016, Mead2020a, Mead2021b}, \BCEmu\citep{Schneider2019Baryonification, Giri2021Baryon}, \Bacco \citep{Arico:2021:Bacco}, and many more. Multiple weak lensing-based analyses have been undertaken using one of these models \citep[\eg][]{Chen:2023:DESY3, Arico:2023:DESY3, Grandis:2024:XrayLensing, Garcia:2024:smallscale, Bigwood:2024:BaryonsWLkSZ, Pandey:2025:Baryons, Dalal:2025:Baryons}. Of these works, \citet{Bigwood:2024:BaryonsWLkSZ} considered a variety of models and performed a systematic comparison of the corresponding results when applied to DES Y3 data. Our analysis here is also motivated by the findings of \citet{Bigwood:2024:BaryonsWLkSZ}; namely, that the data prefer a strong ($\approx 25\%$) suppression in the matter power spectrum.

The above models (\BCEmu, \Bacco, \Hmx) differ significantly in their exact model parameterization, implementations, and assumed astrophysical (and cosmological) priors. We direct the interested reader to the corresponding references above for more details on these models. To enable more normalized comparisons across different choices, we build a single pipeline that predicts the baryon suppression using a halo-model approach \citep[\eg][]{Cooray:2002:HaloModel, Asgari:2023:HaloModel} with the halo density profiles assumed by each of these models (\HMx, \BCEmu, and \Bacco). These predictions are derived using the \textsc{BaryonForge}\footnote{\url{https://github.com/DhayaaAnbajagane/BaryonForge}} codebase \citep{Anbajagane:2024:Baryonification}, where we have implemented all profiles from these models. We then build an emulator for the baryon suppression that spans both astrophysical nuisance parameters and cosmology parameters. All halo-model calculations are implemented using the \textsc{Core Cosmology Library} \citep[CCL,][]{Chisari2019CCL}. The exact modeling choices are provided in Appendix \ref{appx:BaryonModel}. In all results to follow, we refer to the \BaryonForge-based calculation of \BCEmu, \Bacco, and \HMx as BFG-\BCEmu, BFG-\Bacco and BFG-\HMx, respectively.

\begin{figure}
    \centering
    \includegraphics[width=1\columnwidth]{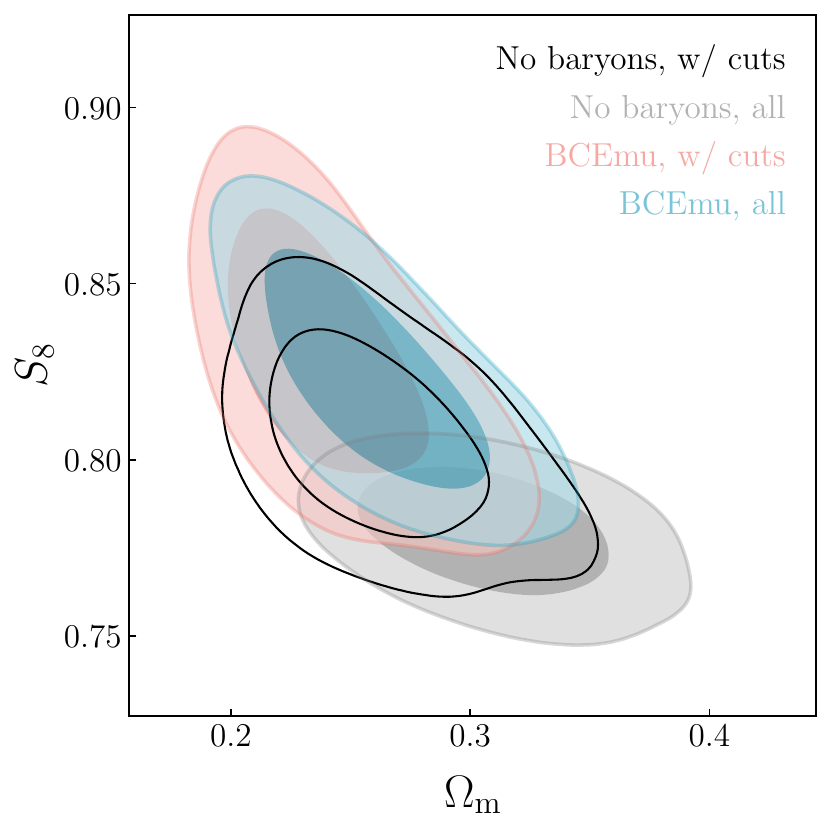}
    \caption{The \LCDM constraints from (i) using our fiducial model (``No baryons'') or one that includes baryon power suppression (``\BCEmu'') and (ii) using a data vector with all scales or with scale cuts. The constraints from including the baryon modeling (with and without scale cuts) is consistent with those from the fiducial model with scale cuts. The results indicate that after scale cuts are applied, the residual bias is $0.3\sigma$ in the $\Seight$-$\Om$ plane. Using all scales with an explicit baryon model has a 7\% higher FoM compared to using scale cuts with no baryon model. See Section \ref{sec:Baryons} for more discussion.}
    \label{fig:BaryonScaleCuts}
\end{figure}

\begin{figure*}
    \centering
    \includegraphics[width=2\columnwidth]{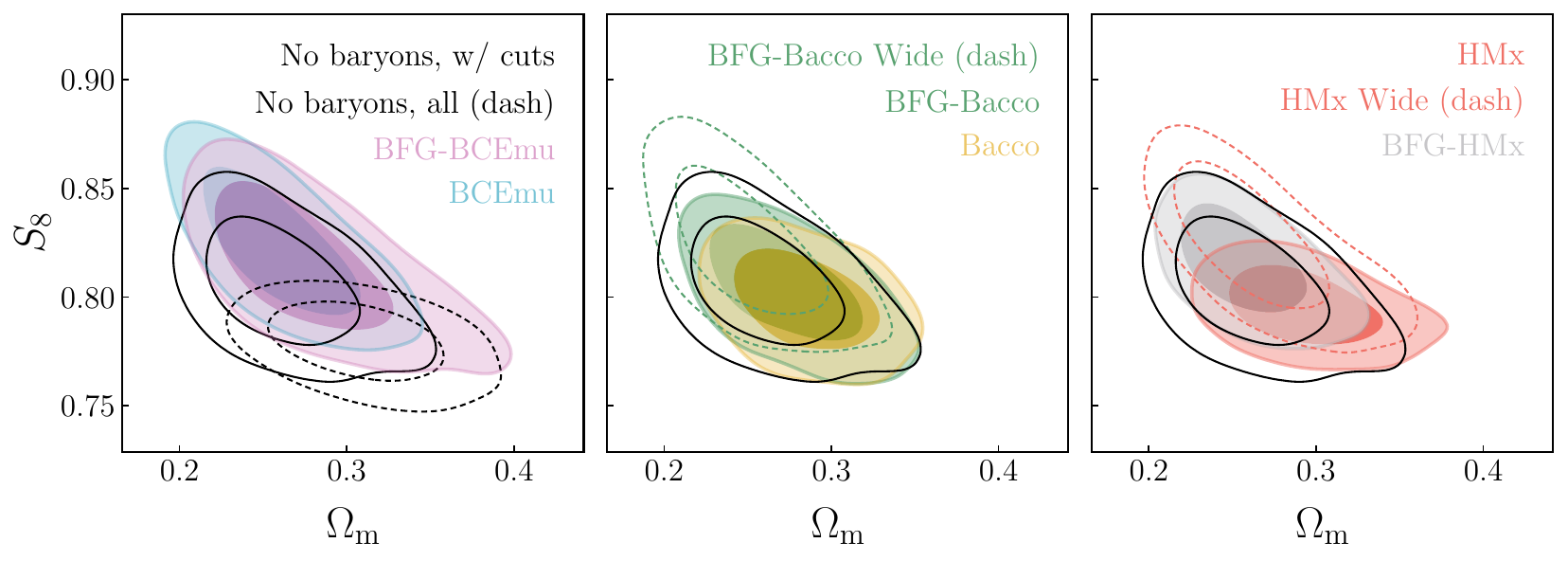}
    \caption{The \LCDM constraints from using all available measurements/scales and marginalizing over the impact of baryons on the power spectrum, where this impact is modelled via different formalisms (different panels). We use publicly available modules and corresponding halo-model approaches evaluated under the common framework of the \textsc{BaryonForge} pipeline. The resulting predictions for the power-spectrum suppression are shown in Figure \ref{fig:PkSuppression}. See Section \ref{sec:Baryons} for details on the different variants. We find that different baryon modeling frameworks can achieve concordance in cosmology constraints once the models marginalize over sufficiently wide astrophysical priors.}
    \label{fig:BaryonConstraints}
\end{figure*}

To reiterate, our cosmology analysis pipeline now includes an additional step: the matter power spectrum is rescaled by the predicted suppression from a given baryon model. As mentioned previously, the analysis in this section uses the entire cosmic shear data vector from all three datasets (NGC, SGC, DES Y3) and does not apply scale cuts to any of them. All other parts of our cosmology inference pipeline remain identical to the fiducial analysis of Section \ref{sec:results:LCDM}. 

\subsubsection{Baryon suppression of the matter power spectrum}

Figure \ref{fig:PkSuppression} presents the baryon suppression of the matter power spectrum predicted by each model. This is shown for our fiducial analysis using the DECam 13k data, and for the extended analysis which adds BAO and SNe information to the WL data. Constraints from the two different data combinations are consistent with each other, so we focus our discussion on the former setup. The astrophysical priors for the different models are listed in Table \ref{tab:baryonparams}. For \HMx, we use the recommended prior $T_{\rm agn} \in [10^{7.6}, 10^{8.0}]K$ as in \citet{Mead2020a}. The results for the different models in Figure \ref{fig:PkSuppression} show many features. First, our \BaryonForge-based replication of \BCEmu and \Bacco produces similar results to the published emulators. In Figure \ref{fig:Testfixcosmo} and Appendix \ref{appx:BaryonModel}, we show the remaining differences between BFG-\BCEmu and \BCEmu can be accounted for by the fact that the latter model was trained within a narrower cosmology prior.\footnote{Both the \BCEmu and \Bacco emulators are trained within a cosmology prior that is narrower than the ones used in lensing analyses. A common solution---which is also the one employed in this work---is to extrapolate these emulators by using predictions from the nearest point in the prior \citep[\eg][]{Arico:2023:DESY3}. This is not required for our \BaryonForge-based models as these models are trained over the same cosmology prior used in the lensing analysis.} Differences between BFG-\Hmx and \Hmx are expected as the latter varies the single ``effective'' parameter, $T_{\rm agn}$, whereas the former varies 10 parameters. See Appendix \ref{appx:BaryonModel} for more details.

Figure \ref{fig:PkSuppression} reproduces a known result --- for the same dataset, the \Bacco and \HMx models predict shallower suppression than \BCEmu \citep[\eg][]{Chen:2023:DESY3, Arico:2023:DESY3, Garcia:2024:smallscale, Bigwood:2024:BaryonsWLkSZ}. The work of \citet{Bigwood:2024:BaryonsWLkSZ} shows the difference between \HMx and \BCEmu can be alleviated by widening the priors on the former, and we confirm the same by changing the priors to $T_{\rm agn} \in [10^7, 10^9] K$. We now show a similar scenario is true for the \Bacco modeling, and this is done by retraining our BFG-\Bacco model with a broader prior on the mass-scaling parameter, $M_c$.\footnote{Existing analyses of \Bacco on DES Y3 data show this parameter's posterior is limited by the upper bound of the prior \citep{Arico:2023:DESY3, Chen:2023:DESY3}. That is, the DES Y3 data prefer larger values of $M_c$ than is allowed by the prior. We note that a larger $M_c$ predicts stronger suppression to the matter power spectrum.} See Appendix \ref{appx:BaryonModel} for more details on this variant. The retrained version (``BFG-\Bacco Wide'') prefers a stronger suppression and is now in agreement with \BCEmu.

Therefore, multiple different parameterizations of the baryon suppression produce similar results, provided that the priors on model parameters are sufficiently wide. This consistency across models is a non-trivial result and highlights the robustness of the inferred suppression. All models are in agreement that the power spectrum suppression has an amplitude of $\approx 25\%$, consistent with the DES Y3-based results of \citet{Bigwood:2024:BaryonsWLkSZ}.

\subsubsection{Cosmology with scale cuts and with all scales}

Having established concordance in the baryon suppression constraints from different models, we now turn to the effectiveness of scale cuts in mitigating baryon-driven biases in the inferred cosmology. For simplicity, we only consider the \BCEmu model for this analysis. We quantify the baryon-driven bias by analyzing the full data vector and scale-cut data vector with a ``No baryons'' model and \BCEmu model. Results from the four possible permutations are shown in Figure \ref{fig:BaryonScaleCuts}. The constraints from the full data vector and scale-cut data vector are consistent within $0.1\sigma$ when analyzed with the \BCEmu model, but the shift grows to $0.8\sigma$ if we do not use any baryon modeling. When focusing on the data vector with scale cuts, the cosmology constraints shift by $0.3\sigma$ if we include/exclude \BCEmu from the modeling pipeline. We interpret this shift as the impact of baryons on cosmology estimated using the data vector with scale cuts. The amplitude of the shift is at the level allowed by different lensing analyses \citep[\eg][\citetalias{paper3, paper4}]{Secco2021,Amon2021}. This result is somewhat expected as the baryon suppression assumed by these works (``OWLS AGN'', see Figure \ref{fig:PkSuppresion_Sims}) when determining scale cuts is similar, though with mildly lower amplitude, to our suppression constraints from the data. 
In summary, Figure \ref{fig:BaryonScaleCuts} and its surrounding results indicate that existing scale cuts minimize the impact of baryon effects to a bias that is $\approx 0.3\sigma$ in the $\Seight-\Om$ plane. The Bayesian evidence supports this interpretation, as the evidence ratio between the fiducial and BCEmu models --- under the standard scale cuts --- is $R_{\rm cuts} = 3.15$, indicating only a mild preference for the BCEmu model. However, once we use all scales, we find $R_{\rm nocuts} = 37.1$, which is a strong preference for this model.

We also note that the constraining power from using all scales, with an explicit baryon model, only marginally improves on that from the fiducial result, ``No baryons, w/ cuts''. Table \ref{tab:constraints} shows the FoM improves from 1872 to 1993, which is a 7\% improvement. The additional scales included in the ``No baryons, all'' analysis primarily inform the baryon suppression model and have minimal impact on the cosmology constraints. This is similar to the results of \citet[][see their Table 3]{Bigwood:2024:BaryonsWLkSZ}.

Finally, Figure \ref{fig:BaryonConstraints} presents the constraints on $\Seight$ and $\Om$ when using the full data vector alongside all the baryon suppression models shown in Figure \ref{fig:PkSuppression}. We see a general trend of tighter constraints, and a preference for high $\Om$ and low $\Seight$, when the corresponding baryon suppression predictions of a model are shallower than that found in \BCEmu. For example, the \Bacco and \HMx constraints are generally tighter but limited to lower $\Seight$ and mildly higher $\Om$. The prior-extended models for \Bacco and \HMx (BFG-\Bacco ``Wide'' and \Hmx ``Wide'') show constraints that closely match the \BCEmu ones. On inspecting the \BCEmu contour and the BFG-\BCEmu analog, we also see the former has no preference (relative to the latter) for higher values of $\Om$. Figure \ref{fig:Testfixcosmo} in Appendix \ref{appx:BaryonModel} confirm this is because the former uses narrower priors on cosmology (which is then circumvented through a nearest-neighbor extrapolation). Note also that the cosmology constraints from the various models are all consistent with \textit{Planck} within $1.5\sigma$ to $1.9\sigma$. The IA constraints also do not shift due to the inclusion of small scales and baryon modeling (Figure \ref{fig:IA} and Appendix \ref{appx:IA}). The evidence ratios for each variant, relative to the case where we use all scales but with no baryon model, spans $33 < R < 62$ depending on the model and the choice of prior. In all cases (assuming all scales are used), there is a clear preference for including baryon suppression modelling.

\textbf{In summary,} the analyses above explicitly show that different baryon models can achieve concordance for the suppression of the matter power spectrum and for the associated cosmology constraints. In some cases (for \Bacco and \Hmx), this requires widening the priors on the astrophysical parameters of the baryon models. There is currently no significant benefit to utilizing all scales of the data vector and modeling small scales with an adequately flexible baryon suppression model. The FoM improves by 7\% relative to the fiducial analysis, where the latter uses scale cuts but does not marginalize over any astrophysical nuisance parameters.

We once again note that the above constraints are obtained using the largest weak lensing catalog to date, improving on the existing catalogs by a factor of 3. Even with this additional constraining power, the baryon suppression cannot be sufficiently ``self-calibrated'' by the lensing data alone. Instead, any additional constraining power from lensing on small scales benefits only the baryon suppression constraints. Current work has shown that the path forward is to use external probes, such as the thermal and kinetic Sunyaev Zeldovich effects \citep{Sunyaev1972SZEffect}, to place data-informed priors on the baryon modeling \citep{Bigwood:2024:BaryonsWLkSZ, Pandey:2025:Baryons, Dalal:2025:Baryons}.

\begin{figure*}
    \centering
    \includegraphics[width=1.9\columnwidth]{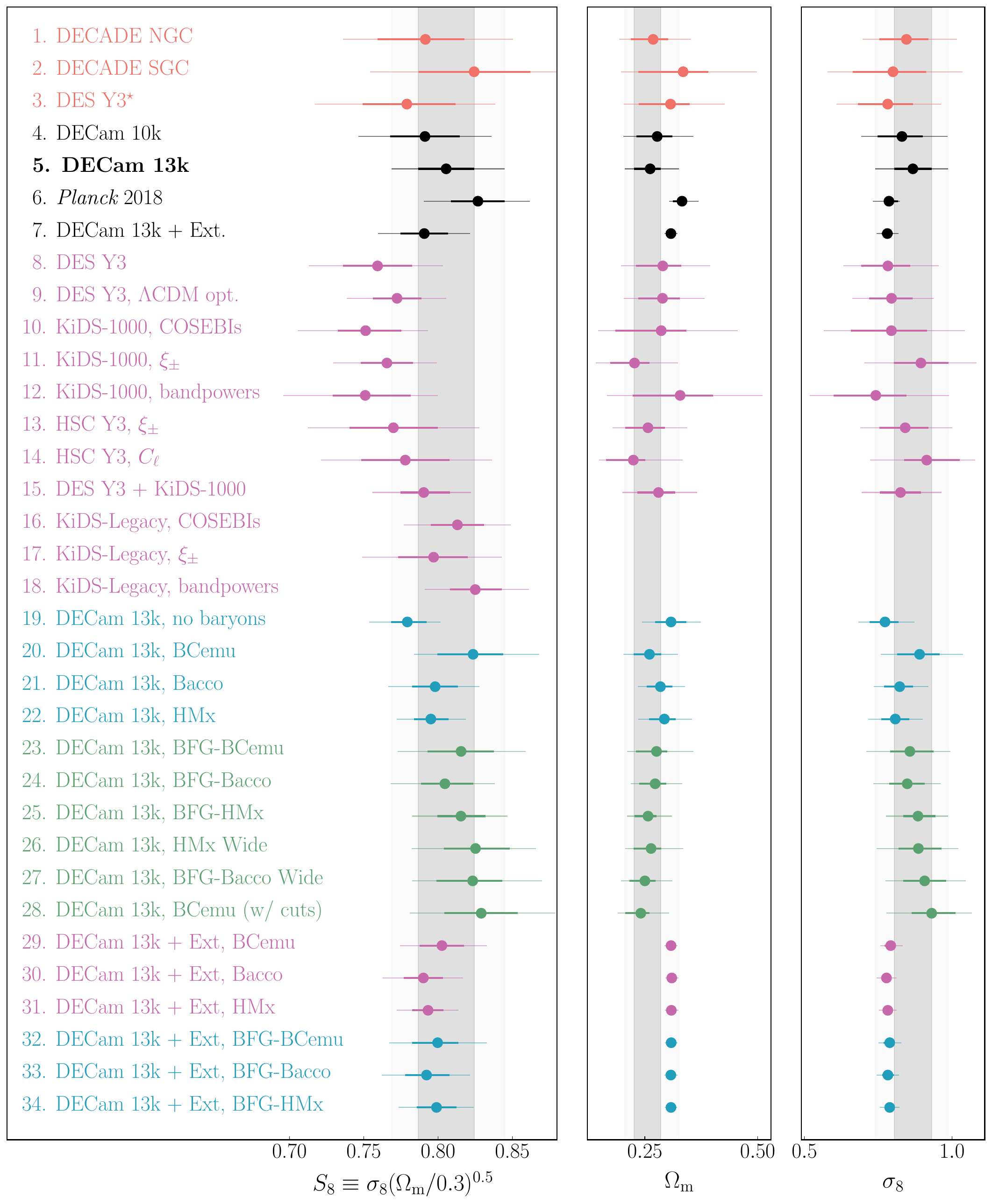}
    \caption{All \LCDM constraints from this work, in comparison to external constraints (lines 8-15) and to variations of different baryon models. The gray bands show the $1\sigma$ and $2\sigma$ regions of the fiducial results (line 5, DECam 13k). The numerical constraints are listed in Table \ref{tab:constraints}. KiDS-Legacy only quotes constraints on $\Seight$.}
    \label{fig:Summary}
\end{figure*}

\section{Summary}\label{sec:summary}

In this work, we present the largest weak lensing analysis to date, using 270 million source galaxies covering 13,000~$\deg^2$ of the sky. This is obtained by combining the \decade NGC, \decade SGC, and DES Y3 datasets which cover 5412~$\deg^2$, 3356~$\deg^2$, and 4143~$\deg^2$, respectively. We use this data to produce constraints on \LCDM, as well as on two models of dynamical dark energy. We then utilize a variety of methods for modeling the impact of baryons on the matter power spectrum, and provide a number of constraints that distinguish between the various approaches.\vspace{10pt}

\noindent Our key results are as follows:\vspace{10pt}

\begin{itemize}
    \item Our fiducial \LCDM constraints, $\Seight = 0.805 \pm 0.019$ and $\Om= 0.262^{+0.023}_{-0.036}$, are within $1.9\sigma$ of the \textit{Planck} 2018 constraints (Figure \ref{fig:LCDM}). Our analysis uses conservative scale cuts and a flexible, five-parameter IA model for each of the three datasets, but still achieves constraints on $\Seight$ that have the same precision as those from \textit{Planck} 2018.
    \vspace{10pt}

    \item The addition of lensing measurements to BAO and SNe data provides noticeable, but still minor, improvements to the constraints on dynamical dark energy models (Figure \ref{fig:darkenergy} and Table \ref{tab:dynamDE}). The combined data is in tension with \LCDM at $3.2\sigma$ for the $w_0w_a$ model, and $3.5\sigma$ for the $w_\phi$ model. 
    \vspace{10pt}

    \item A wide range of baryon modeling approaches/methods --- \BCEmu, \Bacco, \HMx, and alternative halo-model predictions using the \BaryonForge codebase --- all provide consistent constraints for the suppression of the matter power spectrum, and this requires placing sufficiently wide priors on the astrophysics parameters of \HMx and \Bacco (Figure \ref{fig:PkSuppression} and Appendix \ref{appx:BaryonModel}). It also requires using consistent priors for the cosmology parameters (Figure \ref{fig:Testfixcosmo}). \vspace{10pt}

    \item Our data-driven test for baryonic effects shows that with current scale-cut approaches, the constraints in the $\Seight-\Om$ plane are biased by at most $\approx 0.3\sigma$ due to such effects, even with the increased precision of our combined data (Figure \ref{fig:BaryonScaleCuts}). The scale cuts in both our work and past work are derived assuming the ``OWLS AGN'' model for power suppression. This model is fairly consistent with the predicted suppression from our data (Figure \ref{fig:PkSuppression}). \vspace{10pt}
    
    \item For a lensing-only analysis, it is not necessarily advantageous to use all scales alongside explicit baryon modeling as this improves the figure-of-merit in $\Seight, \Om$ by only 7\%, relative to the fiducial analysis (Table \ref{tab:constraints}).
    The new information provided by the small scales constrains only the additional degrees of freedom in the baryon model. External probes are necessary for optimally using small-scale lensing data. This statement assumes baryons are the only statistically significant systematic on small scales.\vspace{10pt}
\end{itemize}

The current precision of our lensing analyses is limited primarily by two factors: (i) uncertainty in the baryon suppression or alternatively, the use of scale-cuts to remove baryon sensitivity, and (ii) uncertainty in the intrinsic alignment signals of source galaxies. The former can be improved through multi-probe analyses \citep[\eg][]{Bigwood:2024:BaryonsWLkSZ, Pandey:2025:Baryons, Dalal:2025:Baryons} while the latter can be improved either with direct measurements of IA from spectroscopic datasets \citep[\eg][]{Samuroff2022, Georgiou:2025:IA, Siegel:2025:IA} or by utilizing more stringent selections on the source galaxy sample \citep{McCullough:2024:BlueShear}. The uncertainties from redshift calibration are negligible relative to the other two; though, we note this is also because \decade and DES Y3 data are shallower than the data from DES Y6 \citep{Yamamoto2025} and the upcoming LSST Y1 release. The calibration uncertainties will grow larger as we incorporate fainter objects into our source galaxy samples. In summary, while our analysis --- like other lensing analyses --- is limited by modeling uncertainties, there is a clear path to different data-driven approaches that can reduce such uncertainties in the near future.

{\setlength{\tabcolsep}{11pt}
\begin{table*}
    \centering
    \begin{tabular}{ccccccc}
    \hline \rule{0pt}{10pt}
    Run & $\Seight$ & $\Om$ & $\sigma_8$ & $\chi^2/N_{\rm dof}$ & $p$ & FoM$_{\Seight \Om}$\\
    \hline \rule{0pt}{10pt}
    DECADE NGC & $0.791^{+0.026}_{-0.032}$ & $0.268^{+0.033}_{-0.050}$ & $0.845^{+0.075}_{-0.092}$ & 266.5/220 & 0.017 & 861 \\[3pt]
    DECADE SGC & $0.824\pm 0.037$ & $0.335^{+0.056}_{-0.099}$ & $0.80^{+0.11}_{-0.14}$ & 244.9/234 & 0.299 & 447 \\[3pt]
    DES Y3$^\star$ & $0.779\pm 0.031$ & $0.307^{+0.043}_{-0.071}$ & $0.782^{+0.086}_{-0.10}$ & 238.6/227 & 0.286 & 615 \\[3pt]
    DECam 10k & $0.791\pm 0.023$ & $0.277^{+0.034}_{-0.046}$ & $0.830^{+0.071}_{-0.082}$ & 502.9/447 & 0.034 & 1240 \\[3pt]
    \textbf{DECam 13k} & $\boldsymbol{0.805\pm 0.019}$ & $\boldsymbol{0.262^{+0.023}_{-0.036}}$ & $\boldsymbol{0.867\pm 0.063}$ & \textbf{774.3/681} & \textbf{0.007} & \textbf{1872} \\[3pt]
    $\textit{Planck}$ 2018 & $0.827\pm 0.018$ & $0.332^{+0.010}_{-0.020}$ & $0.786^{+0.030}_{-0.012}$ & --- & --- & 3252 \\[3pt]
    DECam 13k + Ext. & $0.791\pm 0.016$ & $0.3078\pm 0.0072$ & $0.781\pm 0.019$ & 2444.4/2429 & 0.409 & 8739 \\[3pt]
    \hline \rule{0pt}{10pt}
    DES Y3 & $0.759\pm 0.023$ & $0.290^{+0.041}_{-0.060}$ & $0.783^{+0.075}_{-0.091}$ & 239.9/220 & 0.170 & 926 \\[3pt]
    DES Y3, $\Lambda$CDM opt. & $0.772\pm 0.017$ & $0.289^{+0.039}_{-0.054}$ & $0.795\pm 0.073$ & 285.7/268 & 0.219 & 1362 \\[3pt]
    KiDS-1000, COSEBIs & $0.751^{+0.024}_{-0.019}$ & $0.286^{+0.056}_{-0.10}$ & $0.79^{+0.12}_{-0.14}$ & 82.2/70 & 0.161 & 650 \\[3pt]
    KiDS-1000, $\xi_\pm$ & $0.766\pm 0.018$ & $0.227^{+0.033}_{-0.053}$ & $0.894\pm 0.095$ & 152.1/115 & 0.013 & 1165 \\[3pt]
    KiDS-1000, bandpowers & $0.751^{+0.031}_{-0.022}$ & $0.328^{+0.072}_{-0.10}$ & $0.74^{+0.10}_{-0.14}$ & 260.3/220 & 0.034 & 588 \\[3pt]
    HSC Y3, $\xi_\pm$ & $0.770\pm 0.030$ & $0.257^{+0.037}_{-0.050}$ & $0.841^{+0.078}_{-0.087}$ & 150.0/140 & 0.266 & 786 \\[3pt]
    HSC Y3, $C_{\ell}$ & $0.778\pm 0.030$ & $0.225^{+0.027}_{-0.061}$ & $0.914^{+0.11}_{-0.077}$ & 58.5/60 & 0.531 & 681 \\[3pt]
    DES Y3 + KiDS-1000 & $0.790^{+0.018}_{-0.016}$ & $0.280^{+0.037}_{-0.047}$ & $0.825\pm 0.069$ & 378.0/348 & 0.129 & 1415 \\[3pt]
    KiDS-Legacy, COSEBIs & $0.813^{+0.018}_{-0.018}$ & --- & --- & 127.8/120.5 & 0.307 & --- \\[3pt]
    KiDS-Legacy, $\xi_\pm$ & $0.825^{+0.018}_{-0.017}$ & --- & --- & 413.1/351.5 & 0.013 & --- \\[3pt]
    KiDS-Legacy, bandpowers & $0.797^{+0.023}_{-0.024}$ & --- & --- & 151.0/162.5 & 0.731 & --- \\[3pt]
    \hline \rule{0pt}{10pt}
    13k, no baryons & $0.779^{+0.013}_{-0.011}$ & $0.308\pm 0.034$ & $0.773^{+0.046}_{-0.052}$ & 1289.1/1200 & 0.037 & 2641 \\[3pt]
    13k, BCemu & $0.824^{+0.020}_{-0.024}$ & $0.260^{+0.026}_{-0.035}$ & $0.890^{+0.068}_{-0.076}$ & 1266.1/1200 & 0.090 & 1993 \\[3pt]
    13k, Bacco & $0.798\pm 0.015$ & $0.285^{+0.026}_{-0.030}$ & $0.823^{+0.045}_{-0.052}$ & 1270.5/1200 & 0.077 & 2522 \\[3pt]
    13k, HMx & $0.795\pm 0.012$ & $0.293^{+0.025}_{-0.034}$ & $0.808\pm 0.048$ & 1270.2/1200 & 0.078 & 2957 \\[3pt]
    13k, BFG-BCemu & $0.815\pm 0.022$ & $0.276^{+0.024}_{-0.045}$ & $0.857^{+0.081}_{-0.066}$ & 1269.9/1200 & 0.079 & 1529 \\[3pt]
    13k, BFG-Bacco & $0.805^{+0.019}_{-0.016}$ & $0.273^{+0.025}_{-0.035}$ & $0.848\pm 0.059$ & 1266.9/1200 & 0.088 & 2287 \\[3pt]
    13k, BFG-HMx & $0.815\pm 0.016$ & $0.257^{+0.019}_{-0.030}$ & $0.885^{+0.060}_{-0.050}$ & 1263.3/1200 & 0.100 & 2803 \\[3pt]
    13k, HMx Wide & $0.825\pm 0.022$ & $0.264^{+0.022}_{-0.038}$ & $0.886^{+0.078}_{-0.067}$ & 1270.7/1200 & 0.076 & 1958 \\[3pt]
    13k, BFG-Bacco Wide & $0.823^{+0.020}_{-0.024}$ & $0.250^{+0.023}_{-0.035}$ & $0.907\pm 0.071$ & 1265.1/1200 & 0.094 & 1878 \\[3pt]
    13k, BCemu (w/ cuts) & $0.829\pm 0.025$ & $0.241^{+0.019}_{-0.034}$ & $0.931^{+0.080}_{-0.068}$ & 768.7/681 & 0.011 & 1682 \\[3pt]
    \hline \rule{0pt}{10pt}
    13k + Ext, BCemu & $0.803\pm 0.015$ & $0.3080\pm 0.0073$ & $0.792^{+0.017}_{-0.020}$ & 2934.1/2948 & 0.569 & 9346 \\[3pt]
    13k + Ext, Bacco & $0.790\pm 0.014$ & $0.3096\pm 0.0073$ & $0.778\pm 0.017$ & 2937.5/2948 & 0.551 & 10283 \\[3pt]
    13k + Ext, HMx & $0.793\pm 0.011$ & $0.3086\pm 0.0074$ & $0.782\pm 0.015$ & 2940.5/2948 & 0.535 & 12872 \\[3pt]
    13k + Ext, BFG-BCemu & $0.800^{+0.014}_{-0.017}$ & $0.3084\pm 0.0073$ & $0.789^{+0.018}_{-0.021}$ & 2935.9/2948 & 0.559 & 8216 \\[3pt]
    13k + Ext, BFG-Bacco & $0.792\pm 0.015$ & $0.3076\pm 0.0073$ & $0.783\pm 0.019$ & 2937.1/2948 & 0.553 & 9457 \\[3pt]
    13k + Ext, BFG-HMx & $0.799\pm 0.013$ & $0.3079\pm 0.0071$ & $0.789\pm 0.017$ & 2934.2/2948 & 0.568 & 11001 \\[3pt]
    \hline
    \end{tabular}
    \caption{Numerical constraints corresponding to Figure \ref{fig:Summary}. We show the mean of the posterior and the relative shifts to the 16\% and 84\% percentile values. The DES Y3 and \textit{Planck} 2018 constraints in the topmost group are our rerun constraints for this work, as presented in \citetalias{paper4}. The constraints are separated into four groups: the \LCDM constraints produced in this work, the public constraints from different surveys, the constraints from using all scales and/or baryon modeling, doing the same but now using BAO and supernovae data as well. The KiDS-Legacy results only quote constraints on $\Seight$.}
    \label{tab:constraints}
\end{table*}
}

This work presents the largest weak lensing analysis to date, spanning 270 million galaxies over 13,000 $\deg^2$ of the sky. The entire survey is constructed using images from a single instrument, the Dark Energy Camera, and the images have all been processed by a single framework, the Dark Energy Survey Data Management (DESDM) pipeline. The combined dataset --- $\approx$ 9,000 $\deg^2$ of \decade data and $\approx$ 4,000 $\deg^2$ of DES data --- spans comparable sky area as is expected from the upcoming LSST Y1 dataset, though with fewer galaxies. As the community awaits the flagship data releases from the next phase of lensing experiments, such as LSST and \textit{Euclid} \citep{Euclid}, this dataset serves as an early preview of a coherent catalog covering one-third of the sky with high fidelity. 

The \decade cosmic shear project has leveraged the weak lensing correlations in this data to extract precise cosmology constraints. However, as the last decade of precision photometric datasets have shown, there is a vast landscape of science --- spanning both astrophysics and cosmology --- that can be accessed with such a dataset, ranging from galaxy cluster profiles \citep[\eg][]{Chang:2019:Splashback, Shin:2019:Splashback, Anbajagane:2024:Shocks, Mpetha:2025:WL} to cross-correlations with other tracers \citep[\eg][]{Ammazzalorso:2020:gammaray, Tilman:2022:tSZxWL, Chang2023, Omori:2023:CMBL, Ferreira:2024:Xray, Pandey:2025:Baryons, LaPosta:2025:XrayWL, Sharma:2026:FRBs} to higher-order statistics of the lensing field \citep[\eg][]{Fluri:2022:KiDS, Gatti:2024:WPH, Jeffrey:2025:Likelihood, Prat:2025:Homology,Cheng:2025:WPH} and many more. In order to better facilitate this wide breadth of science, we publicly release the \decade shear catalogs, data vectors, and associated data products. We hope the \decade dataset provides the community an apt avenue for such work at the dawn of the next generation of wide-field photometric surveys.

\section*{Acknowledgements}

We thank David Herrera, Alice Jacques and Robert Nikutta for their assistance with the public release of the DELVE and \decade data at the Astro Data Lab, and the referee for helping improve the presentation of this paper. DA also thanks Eric Morganson for the inception of the \decade acronym. 

DA is supported by the National Science Foundation (NSF) Graduate Research Fellowship under Grant No.\ DGE 1746045. 
CC is supported by the Henry Luce Foundation and Department of Energy (DOE) grant DE-SC0021949. 
The DECADE project is supported by NSF AST-2108168 and AST-2108169.
The DELVE Survey gratefully acknowledges support from Fermilab LDRD (L2019.011), the NASA {\it Fermi} Guest Investigator Program Cycle 9 (No.\ 91201), and the NSF (AST-2108168, AST-2108169, AST-2307126,  AST-2407526, AST-2407527, AST-2407528). This work was completed in part with resources provided by the University of Chicago’s Research Computing Center. The project that gave rise to these results received the support of a fellowship from "la Caixa" Foundation (ID 100010434). The fellowship code is LCF/BQ/PI23/11970028. C.E.M.-V. is supported by the international Gemini Observatory, a program of NSF NOIRLab, which is managed by the Association of Universities for Research in Astronomy (AURA) under a cooperative agreement with the U.S. National Science Foundation, on behalf of the Gemini partnership of Argentina, Brazil, Canada, Chile, the Republic of Korea, and the United States of America.

Funding for the DES Projects has been provided by the U.S. Department of Energy, the U.S. National Science Foundation, the Ministry of Science and Education of Spain, 
the Science and Technology Facilities Council of the United Kingdom, the Higher Education Funding Council for England, the National Center for Supercomputing 
Applications at the University of Illinois at Urbana-Champaign, the Kavli Institute of Cosmological Physics at the University of Chicago, 
the Center for Cosmology and Astro-Particle Physics at the Ohio State University,
the Mitchell Institute for Fundamental Physics and Astronomy at Texas A\&M University, Financiadora de Estudos e Projetos, 
Funda{\c c}{\~a}o Carlos Chagas Filho de Amparo {\`a} Pesquisa do Estado do Rio de Janeiro, Conselho Nacional de Desenvolvimento Cient{\'i}fico e Tecnol{\'o}gico and 
the Minist{\'e}rio da Ci{\^e}ncia, Tecnologia e Inova{\c c}{\~a}o, the Deutsche Forschungsgemeinschaft and the Collaborating Institutions in the Dark Energy Survey. 

The Collaborating Institutions are Argonne National Laboratory, the University of California at Santa Cruz, the University of Cambridge, Centro de Investigaciones Energ{\'e}ticas, 
Medioambientales y Tecnol{\'o}gicas-Madrid, the University of Chicago, University College London, the DES-Brazil Consortium, the University of Edinburgh, 
the Eidgen{\"o}ssische Technische Hochschule (ETH) Z{\"u}rich, 
Fermi National Accelerator Laboratory, the University of Illinois at Urbana-Champaign, the Institut de Ci{\`e}ncies de l'Espai (IEEC/CSIC), 
the Institut de F{\'i}sica d'Altes Energies, Lawrence Berkeley National Laboratory, the Ludwig-Maximilians Universit{\"a}t M{\"u}nchen and the associated Excellence Cluster Universe, 
the University of Michigan, NSF's NOIRLab, the University of Nottingham, The Ohio State University, the University of Pennsylvania, the University of Portsmouth, 
SLAC National Accelerator Laboratory, Stanford University, the University of Sussex, Texas A\&M University, and the OzDES Membership Consortium.

The DES data management system is supported by the National Science Foundation under Grant Numbers AST-1138766 and AST-1536171.
The DES participants from Spanish institutions are partially supported by MICINN under grants ESP2017-89838, PGC2018-094773, PGC2018-102021, SEV-2016-0588, SEV-2016-0597, and MDM-2015-0509, some of which include ERDF funds from the European Union. IFAE is partially funded by the CERCA program of the Generalitat de Catalunya.
Research leading to these results has received funding from the European Research
Council under the European Union's Seventh Framework Program (FP7/2007-2013) including ERC grant agreements 240672, 291329, and 306478.
We  acknowledge support from the Brazilian Instituto Nacional de Ci\^encia
e Tecnologia (INCT) do e-Universo (CNPq grant 465376/2014-2).

Based in part on observations at Cerro Tololo Inter-American Observatory at NSF's NOIRLab, which is managed by the Association of Universities for Research in Astronomy (AURA) under a cooperative agreement with the National Science Foundation.

This work has made use of data from the European Space Agency (ESA) mission {\it Gaia} (\url{https://www.cosmos.esa.int/gaia}), processed by the {\it Gaia} Data Processing and Analysis Consortium (DPAC, \url{https://www.cosmos.esa.int/web/gaia/dpac/consortium}).
Funding for the DPAC has been provided by national institutions, in particular the institutions participating in the {\it Gaia} Multilateral Agreement.

This paper is based on data collected at the Subaru Telescope and retrieved from the HSC data archive system, which is operated by the Subaru Telescope and Astronomy Data Center (ADC) at NAOJ. Data analysis was in part carried out with the cooperation of Center for Computational Astrophysics (CfCA), NAOJ. We are honored and grateful for the opportunity of observing the Universe from Maunakea, which has the cultural, historical and natural significance in Hawaii. 

This research uses services or data provided by the Astro Data Lab, which is part of the Community Science and Data Center (CSDC) Program of NSF NOIRLab. NOIRLab is operated by the Association of Universities for Research in Astronomy (AURA), Inc. under a cooperative agreement with the U.S. National Science Foundation.

This manuscript has been authored by Fermi Forward Discovery Group, LLC under Contract No.\ 89243024CSC000002 with the U.S. Department of Energy, Office of Science, Office of High Energy Physics.

All analysis in this work was enabled greatly by the following software: \textsc{Pandas} \citep{Mckinney2011pandas}, \textsc{NumPy} \citep{vanderWalt2011Numpy}, \textsc{SciPy} \citep{Virtanen2020Scipy}, \textsc{Matplotlib} \citep{Hunter2007Matplotlib}, and \textsc{Getdist} \citep{Lewis:2019:Getdist}. We have also used
the Astrophysics Data Service (\href{https://ui.adsabs.harvard.edu/}{ADS}) and \href{https://arxiv.org/}{\texttt{arXiv}} preprint repository extensively during this project and the writing of the paper.

\section*{Data Availability}

All catalogs and derived data products (data vectors, redshift distributions, calibrations etc.) for the cosmology analysis are now publicly available through the Noirlab Datalab portal \citep{Fitzpatrick:2014:DataLab, Nikutta:2020:DataLab} as well as through Globus and other avenues. Please visit \url{dhayaaanbajagane.github.io/data_release/decade} for a list of the available dataproducts and their corresponding data access. Our intention is to make all useful products immediately available to the community. Please reach out to DA if a data product of interest to you is not on the above list.

\bibliographystyle{mnras}
\bibliography{References}



\appendix

\begin{table*}
    \centering
    \begin{tabular}{c|c|c|c|c|c|c|c|c|c|c|c|c|c}
     \hline
     & $n$ & $R_{\gamma, 1}$ & $R_{S, 1}$ & $R_{\rm tot, 1}$ & $R_{\gamma, 2}$ & $R_{S, 2}$ & $R_{\rm tot, 2}$ & $n_{\rm eff, C13}$ & $\sigma_{e, {\rm C13}}$ &$n_{\rm eff, H12}$ & $\sigma_{e, {\rm H12}}$ & $\langle \gamma_{1}\rangle$  & $\langle \gamma_{2}\rangle$ \\
     \hline
    Bin 1& 1.312 & 0.854 & -0.010 & 0.843 & 0.855 & -0.011 & 0.844 & 1.167 & 0.234 & 1.174 & 0.234 & -0.00006 & -0.00004  \\
    Bin 2& 1.296 & 0.761 & 0.017 & 0.778 & 0.762 & 0.017 & 0.778 & 1.067 & 0.260 & 1.084 & 0.262 & -0.00009 & 0.00001  \\
    Bin 3& 1.294 & 0.723 & 0.024 & 0.748 & 0.725 & 0.024 & 0.748 & 1.074 & 0.248 & 1.102 & 0.251 & 0.00006 & -0.00010  \\
    Bin 4& 1.292 & 0.594 & 0.031 & 0.625 & 0.595 & 0.033 & 0.628 & 1.030 & 0.284 & 1.089 & 0.292 & 0.00025 & -0.00018  \\
    Full sample& 5.194 & 0.751 & 0.013 & 0.764 & 0.752 & 0.013 & 0.765 & 4.247 & 0.253 & 4.328 & 0.256 & 0.00001 & -0.00006  \\
    \hline
    \end{tabular}
    \caption{Statistics of the \decade SGC dataset. The number density ($n$), different components of the shear response ($R_{\gamma/S/{\rm tot}, 1/2}$), effective number density of source galaxies ($n_{\rm eff}$) and shape noise ($\sigma_{e}$) in the \citet{Heymans2012} and \citet{Chang2013} definitions, and the mean weighted shear ($\langle \gamma_{1,2} \rangle$), all computed for each of the tomographic bins as well as the full non-tomographic sample. The number densities are calculated with an area of $3,\!356 \deg^2$ and are presented in units of 1/arcmin$^2$. The \decade NGC statistics are shown in Table 2 of \citetalias{paper1}.}
    \label{tab:DR32}
\end{table*}

\begin{figure*}
    \centering
    \includegraphics[width=2\columnwidth]{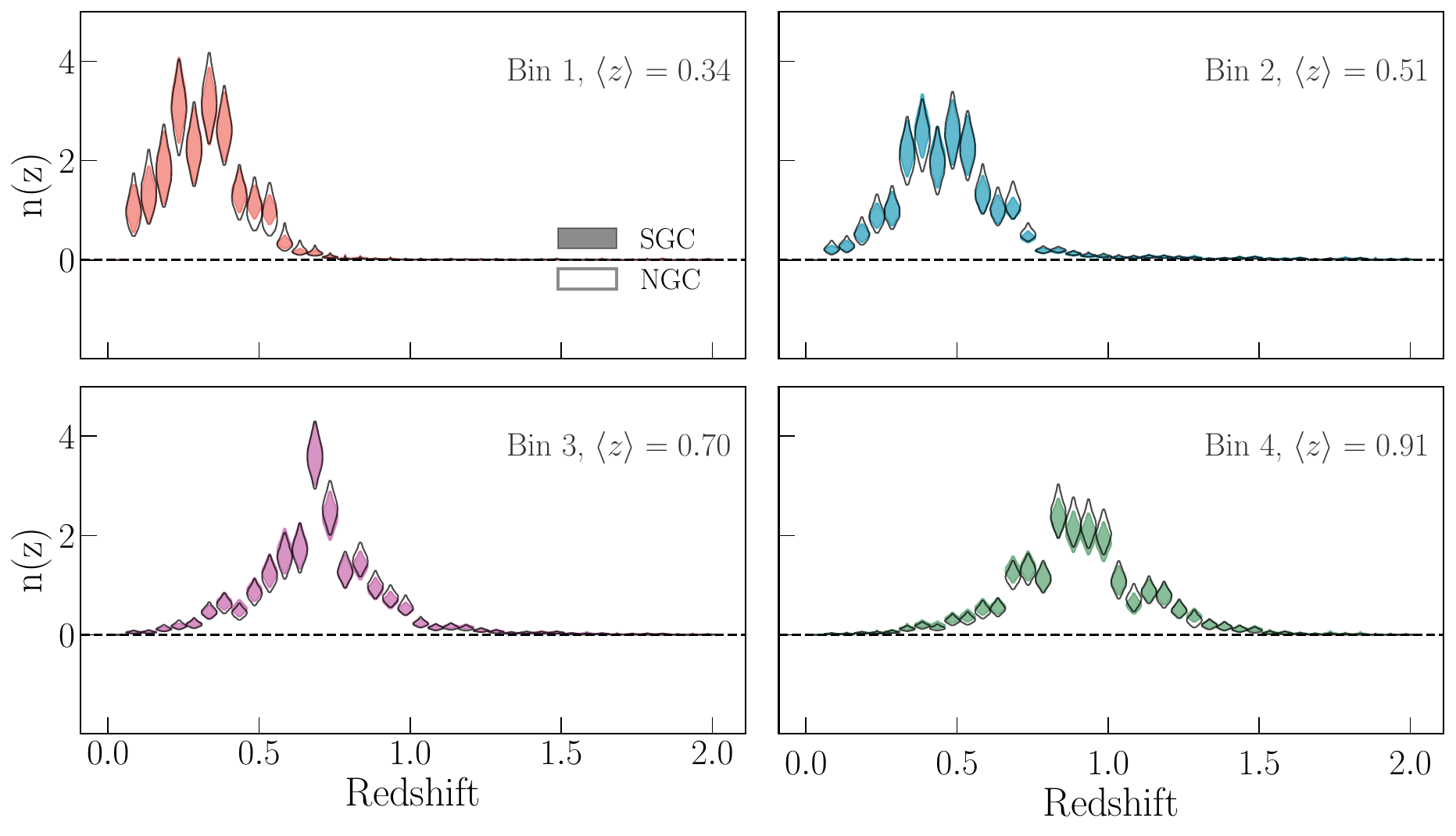}
    \caption{The redshift distribution for each of the four tomographic bins for galaxies in the \decade SGC region. We show the NGC region results for comparison. The distributions are fairly similar across both samples as the dominant uncertainty is from the (small) calibration fields used in the redshift estimation  method \citepalias{paper2}, and these fields are shared across both analyses. See Appendix \ref{appx:SGC} for more details. The mean redshifts for the NGC sample are $\langle z \rangle_i \in \{0.31, 0.50, 0.71, 0.91\}$ for the four tomographic bins.}
    \label{fig:RedshiftsSGC}
\end{figure*}

\section{Data and cosmology constraints from the \decade southern Galactic cap (SGC)} \label{appx:SGC}

The \decade dataset presented in \citetalias{paper1}, \citetalias{paper2}, \citetalias{paper3}, and \citetalias{paper4}, consists of the NGC region of the sky. In this work, we have added an additional dataset covering the SGC region. This dataset spans 3,356 $\deg^2$ (computed using a map of source galaxy counts at \texttt{NSIDE} = 4096) and contains 63 million galaxies. The characteristics of this sample are noted in Table \ref{tab:DR32}. The sky area spanned by the SGC data is presented in Figure \ref{fig:footprint}.

The image and catalog processing follow the exact same approach as \citetalias{paper1}. The one notable addition is a $20\arcdeg$ and $10\arcdeg$ aperture (diameter) circular mask around the Large and Small Magellanic Clouds, respectively, as these regions are particularly crowded. The resulting catalogs have a similar distribution of object properties (fluxes, sizes, \textit{etc.}) as the NGC catalog. The ensemble redshift distributions for the NGC and SGC datasets are shown in Figure \ref{fig:RedshiftsSGC}. They both share similar shapes as the dominant uncertainty on the distributions' morphology is contributions from the (small) datasets used in the calibration --- see the discussion of ``deep field redshift sample'' in \citetalias{paper2} --- and \textit{not} the sample variance in the photometry of the shear catalogs. The former datasets are shared across the NGC and SGC calibrations. 

This statement naturally causes a concern that the redshift calibration uncertainty is correlated across surveys. Our existing pipelines do not jointly produce distributions for both NGC and SGC, and so we are unable to produce correlated calibration priors. Furthermore, we do not have the requisite dataproducts to do the same for DES Y3 dataset. However the analysis of \citet{Cordero:2022:Hyperrank} studied the impact of correlated priors across tomographic bins within a given survey and showed that uncorrelated priors are sufficient in producing unbiased cosmology constraints. In addition, we check that the impact of redshift calibration uncertainties on our final constraints (the DECam 13k combination) is minimal. If we fix the redshift calibration to $\Delta z_i = 0$, the mean of the $\Seight$ marginal posterior changes by $0.05\sigma$ and the posterior width has a fractional change of $5\%$, relative to the fiducial analysis.

We pass the SGC dataset through the same series of tests performed on the NGC dataset, and confirm all tests pass. We provide a brief summary of the tests:
\begin{itemize}
    \item The brighter-fatter effect on the PSFs (the change in PSF size and/or ellipticity with the magnitude of the star) has the same amplitude in the SGC as it did in the NGC (Figure 6 in \citetalias{paper1}), and therefore remains negligible for our analysis.\vspace{10pt}
    
    \item Similarly, the variation of PSF properties with the stellar color is the same across the SGC and NGC (Figure 7 in \citetalias{paper1}), and therefore remains negligible.\vspace{10pt}
    
    \item The correlation of measured galaxy shear, $\gamma^{1,2}$, with other estimated properties (PSF shape, galaxy color, signal-to-noise, PSF size, ratio of galaxy size to PSF size) is statistically consistent with zero in all cases ($<3\sigma$), and the overall variation in the shear is always below $\delta \gamma < 2 \times 10^{-4}$.\vspace{10pt}
    
    \item The Rowe and Tau statistics, which estimate the contamination in lensing due to spatial correlations in the PSF (and PSF errors), have negligible amplitudes. The coefficients for the SGC region --- $\alpha = -0.003 \pm 0.005, \beta = 1.58 \pm 0.11, \eta = 0.28 \pm 1.7$ --- are consistent with our NGC results (Section 4.5 in \citetalias{paper1}) and with DES Y3 \citep{y3-shapecatalog}. Similar to Figure A1 in \citetalias{paper3}, extreme values of PSF contamination have no visible change in the $\Seight - \Om$ contours for the SGC data.\vspace{10pt}
    
    \item The tangential shear around faint (bright) stars is consistent with $p = 0.70$ ($p = 0.13$). \vspace{10pt}
    
    \item The mean shear has no statistically significant correlation with the suite of systematic maps used in \citetalias{paper1} (see their Figure 13). All correlations are consistent with the null signal within $<3\sigma$.\vspace{10pt}
    
    \item The B-modes in the catalog are consistent with the null signal at $p = 0.4$ when using the real-space estimator (Section 4.8 in \citetalias{paper1}). We also check this using a harmonic-space estimator as in Figure A1 of \citetalias{paper3}, and find values of $p = 0.25, 0.11, 0.16, 0.85$ for the four tomographic bins.\vspace{10pt}
    
\end{itemize}
Most of the above tests quantify if there is a large variation in shear with some other property, and we find all variations (in NGC and SGC) are negligible. Similar tests have already been demonstrated in DES Y3 \citep{y3-shapecatalog}. The remaining tests are null tests, which provide a p-value for the measurement being consistent with no signal. We compute the joint probability of the null signal across all three datasets (NGC, SGC, and DES Y3). This can be trivially done as these datasets are all uncorrelated with each other.\footnote{We have three independent p-values, $p_i$. These are distributed as $p_i \in U[0, 1]$. One can show this means $-2\ln p_i \in P(\chi^2_{\rm N = 2})$, i.e. the log p-value is chi-squared distributed with two degrees of freedom \citep{Fisher:1925}, and therefore, the value $-2 \sum_{i=1}^a \ln p_i \in P(\chi^2_{N = {2a}})$. Therefore, the joint likelihood of observing the set of values, $\{p_i\}$ is estimated as $P(\chi^2_{N = 6} > -2\sum_{i=1}^2p_i)$.} We focus on the tangential shear and B-mode tests, which serve as a general metric sensitive to any systematics contaminating galaxy shape estimates. First, the B-modes across all three datasets are consistent with the null signal at $p = 0.38$. Second, the tangential shear around bright (faint) stars is consistent with the null signal at $p = 0.37$ ($p = 0.51$). We therefore find the NGC, SGC, and DES Y3 datasets jointly satisfy these null tests.

The redshift and calibration parameters, estimated using the same procedures as the NGC analysis, are presented in Table \ref{tab:params}. We also carry out the same unblinding procedure for the SGC sample, including the data variations in Figure 6 of \citetalias{paper4} and all split tests from Figure 4 and Figure 6 of \citetalias{paper3}, which pass as well; the constraints on $\Seight$ and $\Om$ from all variations are within $1\sigma$ of the base constraint from fiducial analysis choices. In addition, we ran the split tests on the joint analysis of NGC and SGC, where both catalogs were simultaneously split into halves based on some survey property (depth, seeing, etc.), and constraints were obtained by combining each half with the fiducial DES Y3 dataset (this combination is directly analogous to our ``DECam 13k'' analysis in the main text). The DES Y3 data was not split as we do not have the pipeline to do so. From this analysis, we find the variants are all within $1.5\sigma$ of the fiducial constraints. The cosmology constraints from ``DECam 13k'' show no statistically significant dependence on survey properties. In summary, we have performed the full set of tests described in \citetalias{paper1}, \citetalias{paper2}, \citetalias{paper3}, and \citetalias{paper4}, and find the SGC dataset --- as well as the combination of NCG, SGC, and DES Y3 datasets --- passes all tests.

\begin{figure}
    \centering
    \includegraphics[width=\columnwidth]{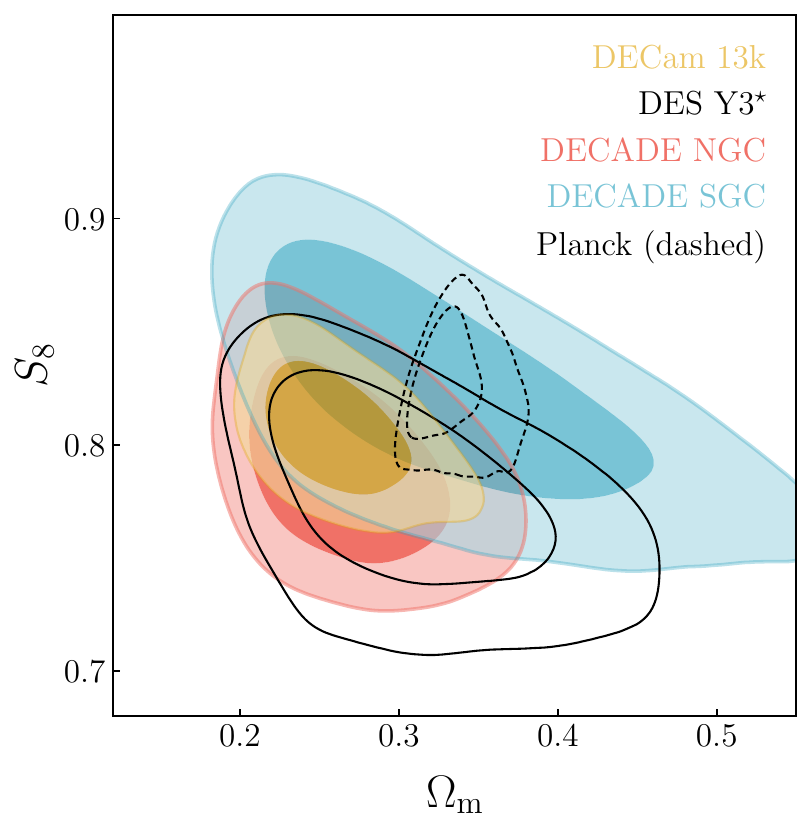}
    \caption{The constraints from the \decade SGC region, compared to \decade NGC, DES Y3, and \textit{Planck}. The three lensing datasets are consistent with each other, within $1\sigma$ in all cases. This motivates the DECam 13k combination presented in the main text and reproduced here as well. The constraints are also listed in Table \ref{tab:constraints}.}
    \label{fig:CosmoSGC}
\end{figure}

Our fiducial cosmology constraints from this sample are presented in Figure \ref{fig:CosmoSGC}. The SGC results are within $1\sigma$ and $0.9\sigma$ of the NGC and DES Y3 results, respectively. All datasets are consistent with one another. The SGC result is also consistent at $<0.01\sigma$ with \textit{Planck}, while the NGC and DES Y3 results are consistent with \textit{Planck} within $1.5\sigma$ and $1.1\sigma$, respectively. We estimate all significances using the Gaussian metric in \textsc{tensiometer}, as is done in the main text. In summary, the SGC dataset has been processed and validated through the same methods in \citetalias{paper1}, \citetalias{paper2}, \citetalias{paper3}, and \citetalias{paper4} and its resulting cosmology constraints are consistent with the NGC and DES datasets. As a result, we can combine all three datasets into the DECam 13k analysis presented in the main text.

\section{Baryon modeling and additional results}\label{appx:BaryonModel}

\begin{table}[]
    \centering
    \begin{tabular}{c|c|c}
        Model & param & prior \\
        \hline
        \hline
         & $\log_{10}M_c$ & [11, 15]\\
         & $\mu$ & [0, 2]\\
         & $\theta_{\rm ej}$ & [2, 8]\\
        BFG-\BCEmu & $\gamma$ & [1, 4]\\
         & $\delta$ & [3, 11]\\
         & $\eta$ & [0.05, 0.4]\\
         & $\delta \eta$ & [0.05, 0.4]\\
         \hline
         & $\log_{10}M_c$ & [11, 15]\\
         & $\log_{10} \eta$ & [-0.7, 0.7]\\
         & $\log_{10} \beta$ & [-1, 0.7]\\
        BFG-\Bacco & $\log_{10} M_{1, 0}$ & [9, 13]\\
         & $\log_{10}\theta_{\rm out}$ & [0, 0.5]\\
         & $\log_{10}\theta_{\rm inn}$ & [-2, -0.5]\\
         & $\log_{10}M_{\rm inn}$ & [9, 13.5]\\
         \hline
         & $A_{\star, 0}$ & [0.01, 0.1]\\
         & $A_{\star, 1}$ & [-0.03, 0.03]\\
         & $M_{\star, 0}$ & [9, 13]\\
         & $M_{\star, 1}$ & [-0.3, 0.3]\\
         BFG-\Hmx& $\eta$ & [-2, 0.1]\\
         & $\epsilon_{1, 0}$ & [-1, 1]\\
         & $\epsilon_{1, 1}$ & [-0.1, 0.1]\\
         & $\Gamma_0$ & [1.05, 5]\\
         & $\Gamma_1$ & [-0.1, 0.1]\\
         & $\log_{10} M_0$ & [11, 16]\\
         \hline
    \end{tabular}
    \caption{The priors on the astrophysics parameters for each model. All parameters are given uniform priors within the range [lower, upper] as denoted in this table. For the parameter definitions and associated priors, see Table 1 in \citet{Giri2021Baryon} for \BCEmu, Section 3.2 in \citet{Arico:2021:Bacco} for \Bacco, and Table 2 in \citet{Mead2020a} for \HMx. For BFG-\Bacco, we also consider two variants: (i) the ``Wide'' variant where we use the updated prior, $\log_{10}M_c \in [11, 17]$, and (ii) the ``Extr'' variant where all parameters have their priors widened by a factor of $2$.}
    \label{tab:baryonparams}
\end{table}

We now detail our modeling approach for the baryon suppression of the matter power spectrum. We use the philosophy of \citet{Mead2021b}, \citet{Pandey:2025:Baryons}, etc. by employing a halo-model formalism for computing the power spectrum $P_{\rm dmo}(k)$ and $P_{\rm bary}(k)$ \citep[see][for reviews]{Cooray:2002:HaloModel, Asgari:2023:HaloModel}. Here ``dmo'' is the dark matter-only model and ``bary'' is the combined dark matter and baryon model. The density profiles in the former comprise a simple Navarro-Frenk-White (NFW) profile \citep{Navarro1997NFWProfile} while those of the latter include a number of components (gas, stars, etc.) that we discuss further below. We can then predict the corresponding power spectrum suppression through the ratio, $S(k) = P_{\rm bary}(k) / P_{\rm dmo}(k)$.

There are a variety of models for the combined dark matter and baryon density profile of the halo. In this work, we use the implementation of profiles in the \BaryonForge codebase \citep{Anbajagane:2024:Baryonification}, which provides all profiles used in the models from \BCEmu \citep{Schneider2019Baryonification, Giri2021Baryon}, \Bacco \citep{Arico:2021:Bacco}, and \HMx \citep{Mead2021b}. We pass these profiles through the halo-model calculator in the \textsc{Core Cosmology Library} \citep[CCL,][]{Chisari2018BaryonsPk} to obtain predictions for the different power spectra. Our halo model calculation uses the halo mass function and halo bias relation from \citet{Tinker:2010:Halobias}. We consider all masses from $10^{10} < M_{\rm 200c} / \msun < 10^{16}$ in the halo-model integral, and utilize the additive correction of \citet{Cacciato:2012:HaloModel} to account for all halos below this mass scale. The rest of the formalism follows the halo-model approach described in  \citet[][and references therein]{Pandey:2025:Baryons}, and we do not replicate the description here.

We do not detail here the exact profiles used for each of the models. We refer readers to \citet{Schneider2019Baryonification, Giri2021Baryon} for the \BCEmu model, \citet{Arico:2021:Bacco} for the \Bacco model, and \citet{Mead2020a} for the \HMx model. The only change made in this work is to the \HMx model, where the ejected gas is now parameterized by an extended profile as in \citet{Schneider:2015:Baryons} and \citet{Arico:2021:Bacco} rather than as an additive constant to the large-scale density field as chosen in \citet{Mead2020a}. 

As mentioned in the main text, we refer to the \BaryonForge-version of these models as BFG-\BCEmu, BFG-\Bacco, and BFG-\Hmx. In all cases, our priors on the baryon parameters in each model follow the same choices as the fiducial models above. The one difference is for \HMx, where the original model is calibrated to a single $T_{\rm agn}$ parameter, whereas BFG-\HMx varies all ten parameters listed in Table 2 of \citet{Mead2020a}. In particular, we use the full list of parameters used for fitting the gas, star, dark matter, and pressure power spectra in that work. Unlike in our approach to \BCEmu and \Bacco, we now vary these ten parameters within a somewhat ad-hoc prior range as we did not have existing results to base our choices on. Our main conclusions are not impacted by this as our \HMx interpretations rely more on the public implementation than our \BaryonForge-based implementation here. The astrophysics priors used for the different models are listed in Table \ref{tab:baryonparams}.

\begin{figure}
    \centering
    \includegraphics[width=\columnwidth]{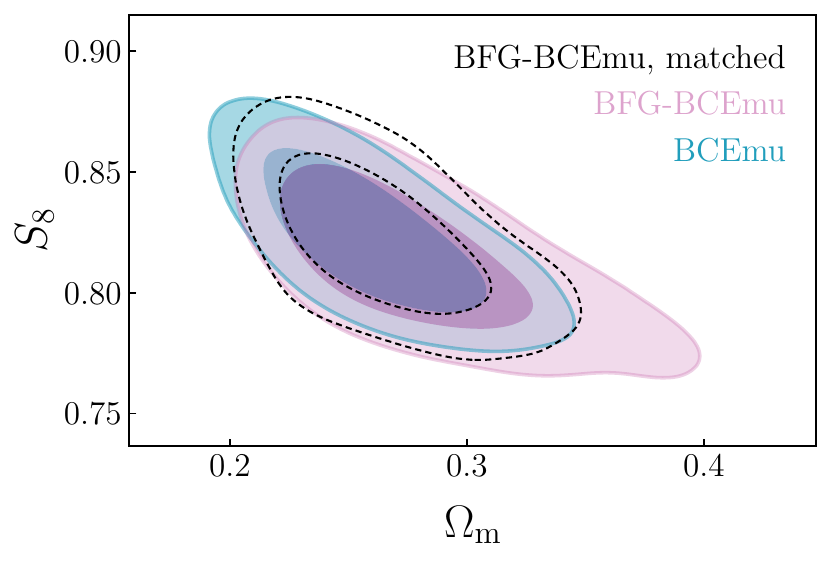}
    \includegraphics[width=\columnwidth]{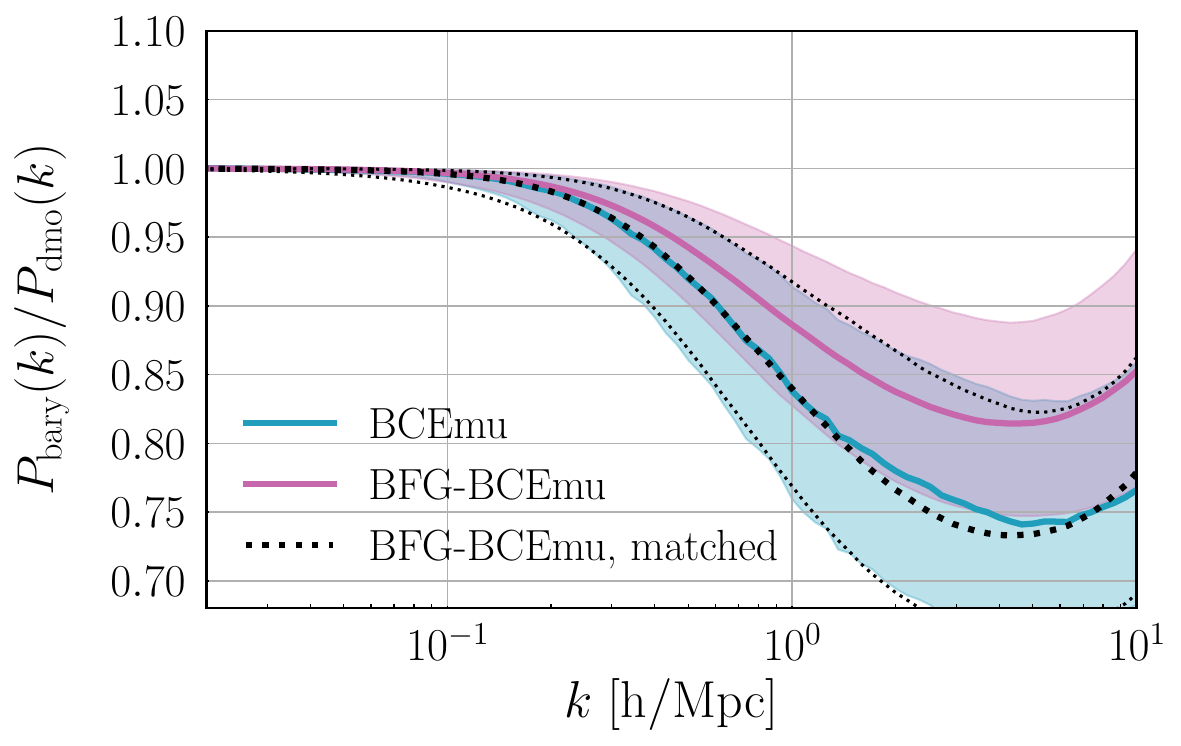}
    \caption{The impact of fixing cosmology priors in the baryon model. The \BCEmu model (blue) fixes all cosmology parameters except the baryon fraction with prior, $f_{\rm b} \in [0.1, 0.25]$. The BFG-\BCEmu model (purple) varies all cosmology parameters within the same prior as Table \ref{tab:params}. If we change the BFG-\BCEmu model to use the fixed priors of \BCEmu (black), then we recover the same constraints on the cosmology (top) and the power spectrum suppression (bottom). The $1\sigma$ posterior on the suppression is denoted in the bands, or the dotted black lines.}
    \label{fig:Testfixcosmo}
\end{figure}

The priors on cosmology parameters are equally important when computing the response. In particular, the baryon modeling is sensitive to variations in $f_b = \Omega_{b}/\Om$ \citep[][see their Figure 2]{Schneider:2020:BaryonEffects}. The public \BCEmu and \Bacco emulators choose a prior on cosmology that is more closely matched to priors from analyses that combine CMB data with other probes. This means the variation in $\Omega_{\rm m}$ is much narrower than the prior used in cosmic shear analyses (Table \ref{tab:params}). We must therefore account for the mismatched prior ranges in some manner. One approach is to simply extrapolate the \BCEmu or \Bacco model using the nearest available point with the prior. While this is a reasonable approach --- and one we take for the fiducial \BCEmu and \Bacco models --- it is still a limitation/inconsistency in the model. In our \BaryonForge-based models of this work, we have control over the prior ranges of the model, and therefore generate $S(k)$ for variations in $\Omega_{\rm b}, \Om, \sigma_8, n_s$ that span the cosmology prior in Table \ref{tab:params}.

Figure \ref{fig:Testfixcosmo} shows the difference in constraints if one uses wider cosmology priors for the baryon model. We start with the \BCEmu and BFG-\BCEmu results of Figure \ref{fig:BaryonConstraints}, and compare them to results of a modified BFG-\BCEmu model where the model's cosmology priors mimic the narrower ones used in \BCEmu. For these priors, we fix all cosmology values except $f_{\rm b}$, which is varied in the range $0.1 < f_{\rm b} < 0.25$. In particular, we fix $\sigma_8 = 0.811$ and $n_s = 0.96$ following the choice made in the \BCEmu emulator \citep[][see their Section 2.2.1]{Giri2021Baryon}. We see that this choice brings the \BCEmu and BFG-\BCEmu models into precise agreement, for both the predicted baryon suppression and for the resulting cosmology constraints. Using a wider cosmology prior, as in BFG-\BCEmu, causes the model to explore slightly larger values of $\Om$. This, in turn, results in a slightly weaker suppression since the baryon fraction is lowered.

\begin{figure}
    \centering
    \includegraphics[width=\columnwidth]{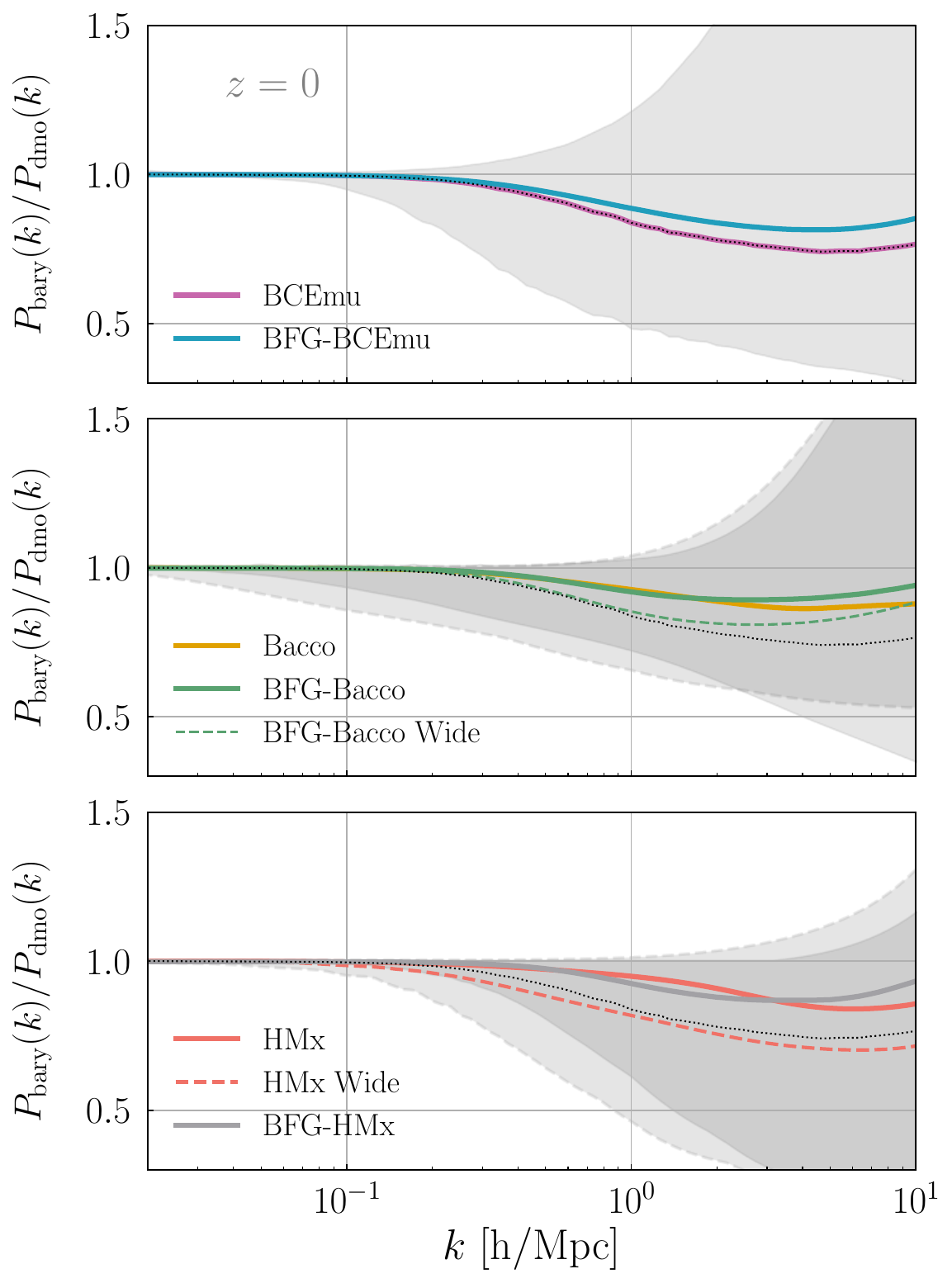}
    \caption{The prior on the baryon suppression (plotted as the maximum and minimum values at each wavenumber, $k$) translated from the prior on the astrophysical and cosmology parameters. In dark gray, we show the prior evaluated on \BCEmu, \Bacco, and \Hmx models, for the top, middle, and bottom panels, respectively. The priors for the BFG-\Bacco Wide and \Hmx Wide analysis are enclosed in dashed gray lines. The dotted black line in each panel is the Fiducial \BCEmu result for the suppression, and is plotted for reference. Even though the \Bacco model has a prior including the \BCEmu suppression, it is unable to reproduce the latter, as the model has more restrictions on the shapes it can predict. Widening the prior to BFG-\Bacco Wide enables a better match.}
    \label{fig:PkSuppresion_Priors}
\end{figure}

\begin{figure}
    \centering
    \includegraphics[width=\columnwidth]{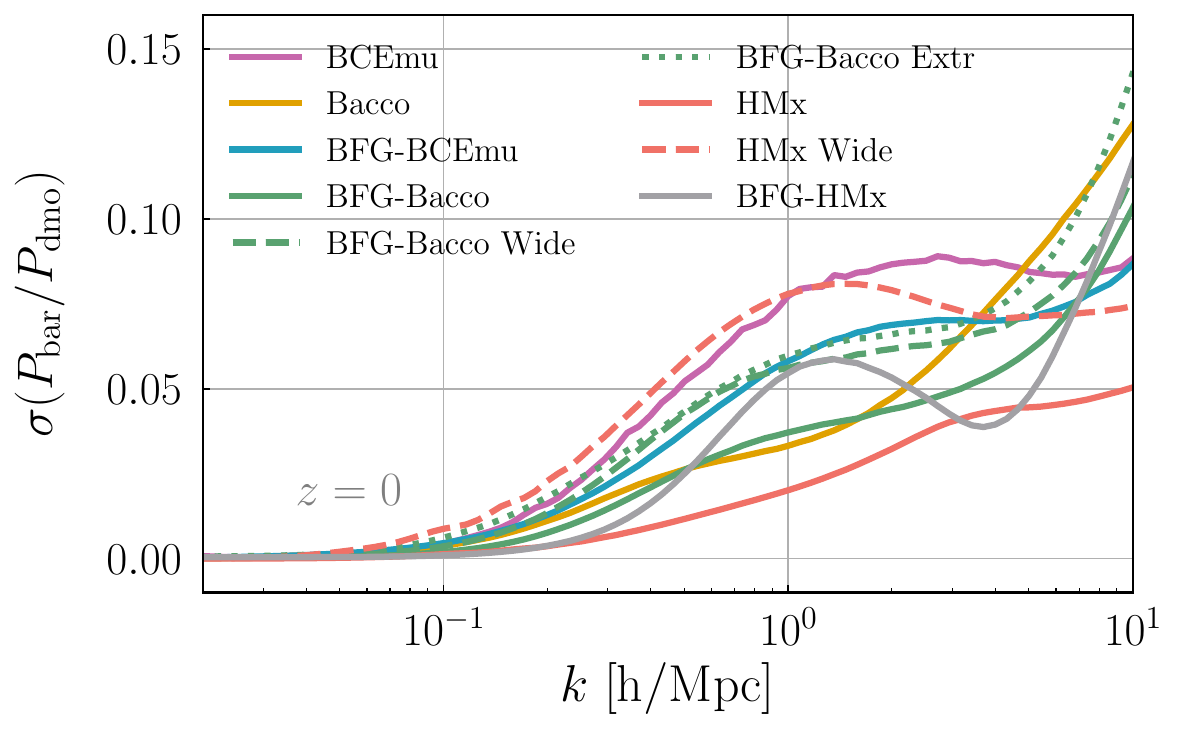}
    \caption{The $1\sigma$ width of the predicted baryon suppression (computed using the 16\% and 84\% percentiles at each wavenumber, $k$) for different models. The models are fairly similar, with the fiducial \Hmx and \Bacco models showing smaller uncertainties as they have more restrictive priors. We show three variants of the \Bacco model: BFG-\Bacco, BFG-\Bacco Wide, and BFG-\Bacco Extr, in order of increasing prior widths. The constraints between Wide and Extr are similar, showing the data is indeed constraining this suppression and is not a purely prior-limited constraint.}
    \label{fig:PkSuppresion_uncert}
\end{figure}

In addition to producing models with wider cosmology priors, we also produce those with some changes to their baryon parameter priors. In particular, the results of \citet{Arico:2023:DESY3} find that the constraints from weak lensing data prefer a value for the mass scale $M_c$\footnote{In \Bacco, this parameter sets the mass scale where the halo has ejected half of its gas mass beyond the associated halo radius \citep{Arico:2021:Bacco}.} that is limited by the upper-bound prior. While the existing \Bacco emulator has the same prior range, we now build our \BaryonForge-based $S(k)$ model for \Bacco with a wider prior in this one parameter (Table \ref{tab:baryonparams}). This enables us to test and show the impact of the parameter prior on the final constraints.

\begin{figure}
    \centering
    \includegraphics[width=\columnwidth]{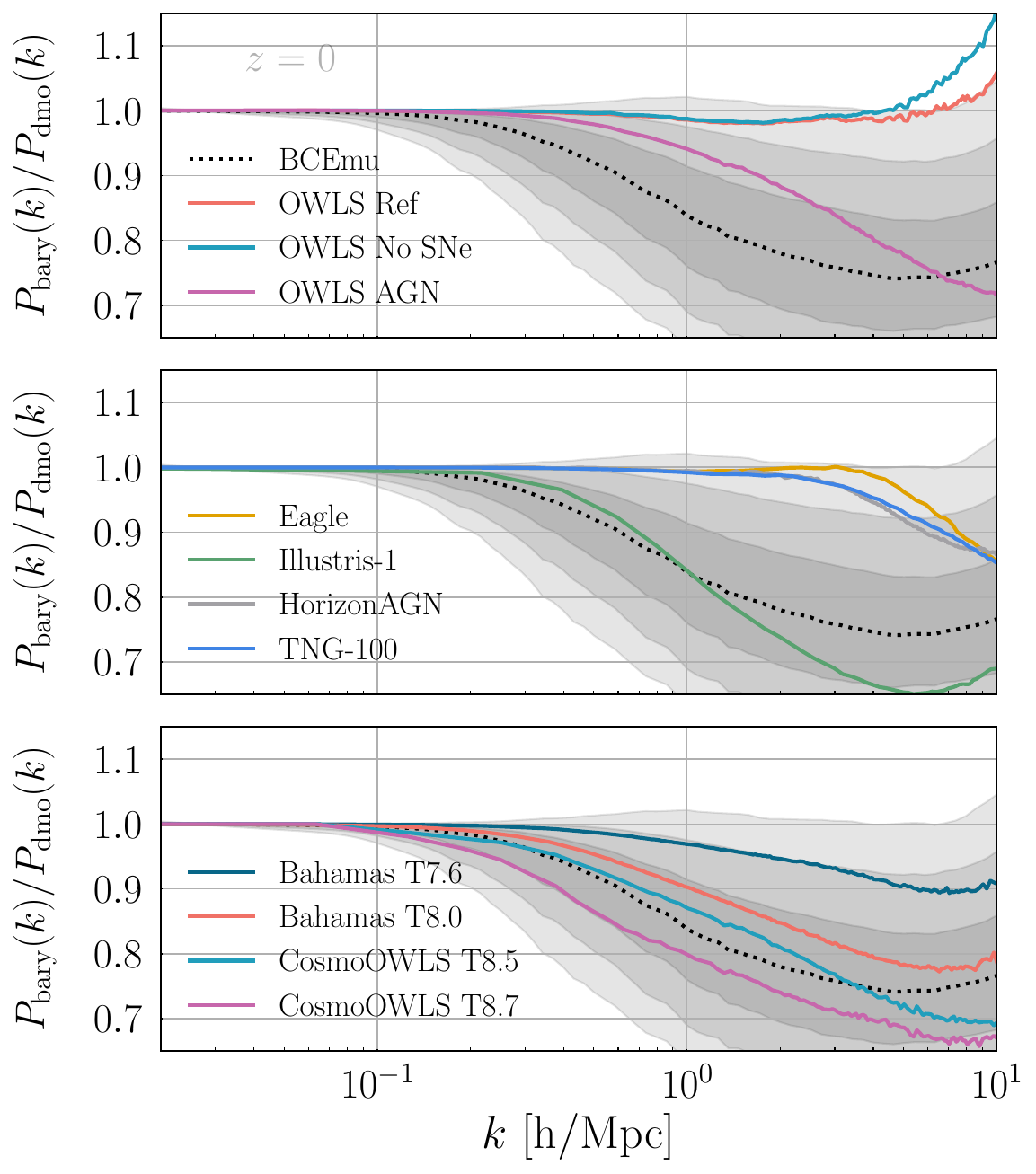}
    \caption{The suppression from the \BCEmu model compared to predictions from different simulations. The gray bands show the 68\%, 95\%, and 99.7\% credible intervals. The constraints from the lensing data are most consistent with the models of OWLS AGN, CosmoOWLS T8.5, and Illustris.}
    \label{fig:PkSuppresion_Sims}
\end{figure}

In all cases, the \BaryonForge-based models are generated at $2^{15} = 32,\!768$ points across the prior of astrophysics and cosmology parameters, and we train a neural network emulator using these points. Our network architecture contains four dense layers. The first three are of size 512, 256, 128, respectively, and the last layer is $N = 100$ to match the dimension of our target suppression (we sample $S(k)$ at 100 values of $k$). After each of the first three dense layers, the inputs go through a \texttt{LeakyReLU} activation with negative slope, $\alpha = 0.1$ before being sent to the next dense layer. We quantify the emulator precision using the commonly used $R^2$ metric,
\begin{equation}
    R_j^2 = 1 - \frac{\sum_i^{\rm N_{\rm test}} (X_{{\rm emu}, i} - X_{{\rm model}, i})^2}{\sum_i^{\rm N_{\rm test}}(X_{{\rm model}, i} - \langle X_{\rm model}\rangle_{i})^2},
\end{equation}
where $X_{{\rm model}, i}$ and $X_{{\rm emu}, i}$ are the model and emulator predictions, respectively, for parameter set, $i$. We evaluate this on $N_{\rm test} = 7,\!000$ test points. The quantity $R^2_j$ is the metric for wavenumber, $k_j$. In practice, we compute $\langle R^2 \rangle_j$ --- \ie, the average over all scales --- as our final test statistic. In all cases, our emulators exhibit $R^2 > 0.99$. We train a separate emulator for each redshift $z \in \{0, 0.5, 1.0, 1.5, 2.0\}$ and linearly interpolate between their results when predicting $S(k)$ for any intermediate redshifts. This follows the approach of \citet{Giri2021Baryon}. The redshift range spanned by our emulator covers the entire range probed by our lensing data, so there is no necessity to extrapolate our model. We have also evaluated the true $S(k)$ model for each sample from our MCMC chains and verified the emulator prediction is accurate (to within $2\%$) across the parameter space spanned by the posteriors.

In Figure \ref{fig:PkSuppresion_Priors}, we translate the priors on the baryon and cosmology parameters into priors on the predicted $S(k)$. Specifically, we show the minimum and maximum suppression values spanned by a model at a given wavenumber. We will denote this range as the ``span'' of the model. We find the span is not a sufficient statistic for validating that the model has an adequately wide prior. For example, the span of the \Bacco model clearly encompasses the \BCEmu constraint (dotted line). However, the final constraints from the model are still shallower in their predicted suppression. The second set of gray bands (enclosed by dashed lines) shows the prior range for the BFG-\Bacco ``Wide'' model, which widens the prior on the BFG-\Bacco model. While the final range of the maximum/minimum suppression at each $k$ is relatively unchanged, the change in the prior still causes differences in the range of shapes of the suppression, which then allows it to better match the \BCEmu result. We highlight this result as a counter-example against using the span of a model to validate its flexibility. 

One may then be concerned that all constraints on the baryon suppression are prior-dominated. However, this is not the case. The suppression discussed in Section \ref{sec:Baryons} are data-informed constraints. To highlight this, we show in Figure \ref{fig:PkSuppresion_uncert} the $1\sigma$ width of the suppression posterior (evaluated using the 16\% and 84\% percentiles). We show a number of models, and also three variants for the BFG-\Bacco model. The ``Wide'' variant increases the prior on $M_c$ as discussed in Section \ref{sec:Baryons}, while the ``Extr'' variant widens the prior by a factor of 2 for all parameters in the model. Notably, the latter constraint has very little difference in its constraining power compared to the former. Thus, the power spectrum suppression --- in the cases where the model prior is sufficiently wide --- is indeed being constrained by data and is not an artifact of the specific prior ranges chosen for the astrophysical nuisance parameters of the model.

Figure \ref{fig:PkSuppresion_Sims} shows our constraints on the suppression in relation to the predictions from various cosmological hydrodynamical simulations. The simulations we compare to are the Overwhelmingly Large Simulations \citep[OWLS,][]{Schaye:2010:OWLS}, Eagle \citep{Schaye:2015:Eagle}, Illustris \citep{Vogelsberger:2014:Illustris}, HorizonAGN \citep{Dubois:2014:HorizonAGN}, IllustrisTNG \citep{Springel:2018:TNG}, Bahamas \citep{McCarthy2017BAHAMAS}, and Cosmo OWLS \citep{LeBrun:2014:COWLS}. In line with existing work \citep[\eg][]{Bigwood:2024:BaryonsWLkSZ, Hadzhiyska:2024:kSZ, Pandey:2025:Baryons, Guachalla:2025:kSZ, Dalal:2025:Baryons}, we find the suppression preferred by the data is best-matched by simulations with stronger AGN feedback prescriptions. In particular, the CosmoOWLS T8.5 model is the closest match, as is the original Illustris model. Figure \ref{fig:PkSuppresion_Sims} also shows that the no-suppression case, $P_{\rm bary}/P_{\rm dmo} = 1$, is excluded at around $3\sigma$.

\begin{figure}
    \centering
    \includegraphics[width=\columnwidth]{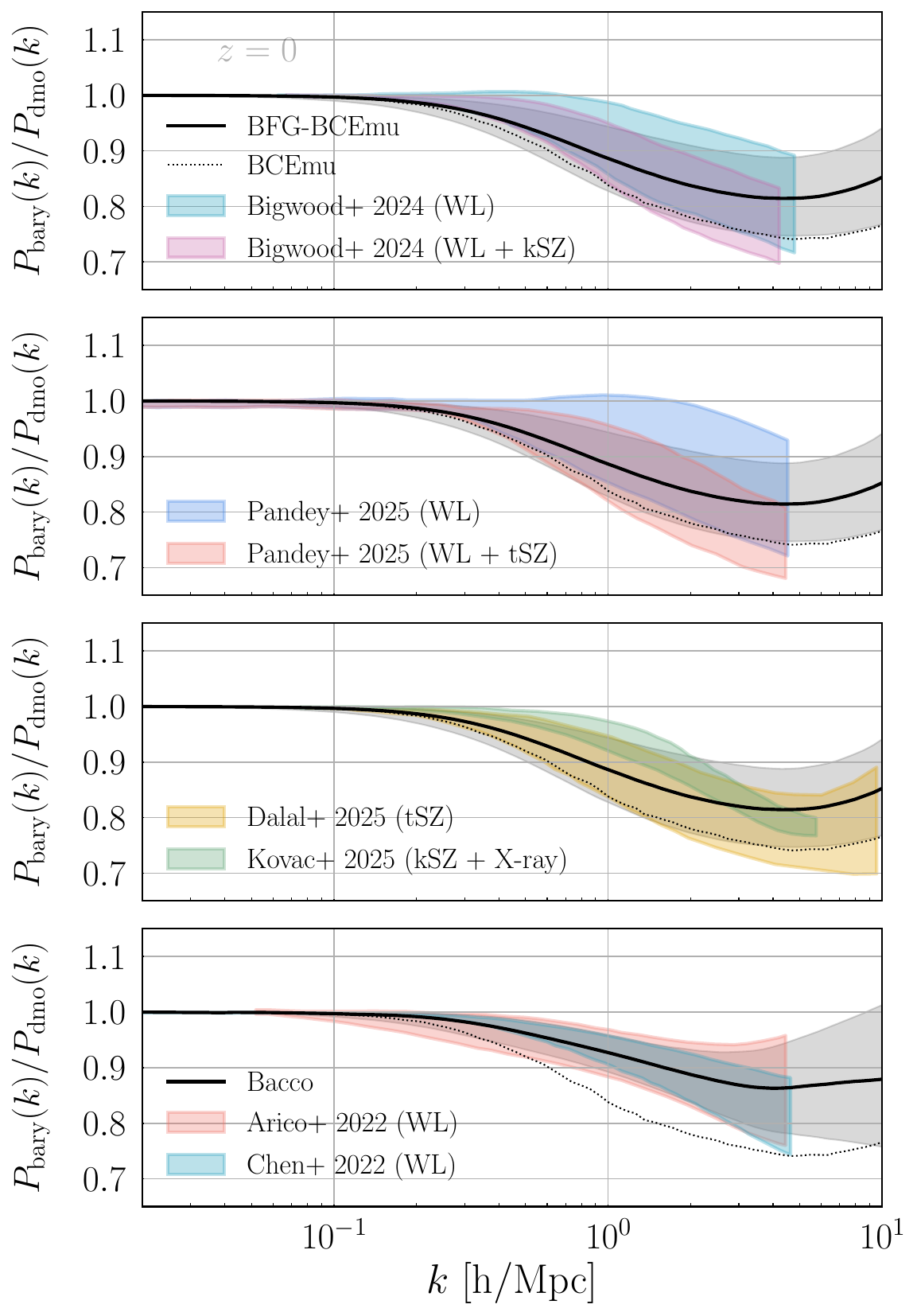}
    \caption{Comparison of our inferred suppression (Figure \ref{fig:PkSuppression}) with other works, which use different models and independent pipelines. The bands denoted the $68\%$ credible interval. The top three panels use the halo profiles defined in \BCEmu, whereas the last panel uses those defined in \Bacco. There is good consistency between the different results. Our \BCEmu result prefers slightly stronger feedback than other works, whereas our BFG-\BCEmu result (which uses a more consistent cosmology prior; see Figure \ref{fig:Testfixcosmo}) is in better agreement with them. The \Bacco constraints have a preferentially shallower amplitude than \BCEmu due to narrower priors (Figure \ref{fig:PkSuppression}). In summary, multiple analyses---crossing different datasets and probes---find a consistent suppression of $\approx 25\%$ in the matter power spectrum.}
    \label{fig:PkSuppresion_Comparison}
\end{figure}

\begin{figure*}
    \centering
    \includegraphics[width=\columnwidth]{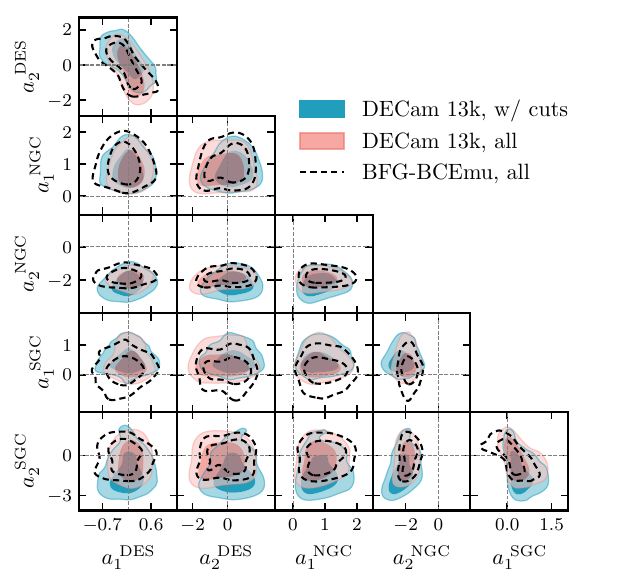}
    \includegraphics[width=\columnwidth]{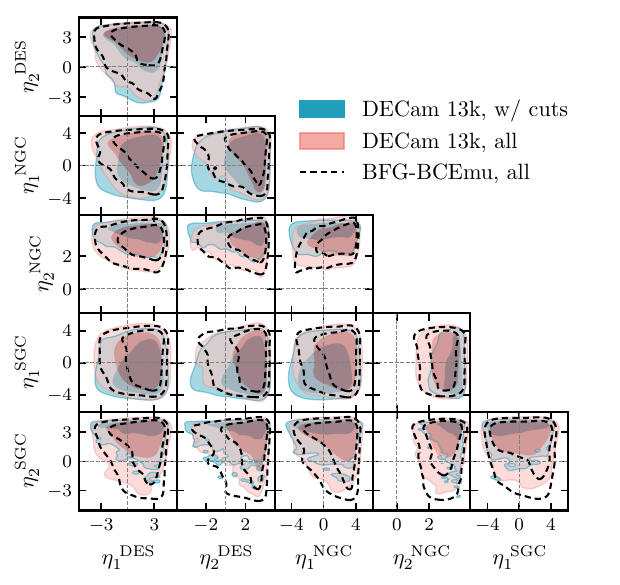}
    \caption{The intrinsic alignment (IA) constraints of different datasets, for the fiducial DECam 13k analysis (blue), using all scales (red), and using all scales with baryon modeling from \BaryonForge (black dashed, see Section \ref{appx:BaryonModel}). The left (right) triangle plot shows the amplitude (redshift scaling) of the IA parameters. We mark $a_i, \eta_i = 0$ in gray dotted lines for reference. The constraints are consistent across the three analysis setups. The NGC dataset shows a strong preference of $a_2 > 0$, which is discussed in detail in \citetalias{paper4} (see their Appendix B1). The DES Y3 and SGC constraints are consistent with each other and show no statistically significant preference for non-zero IA.}
    \label{fig:IA}
\end{figure*}

Finally, Figure \ref{fig:PkSuppresion_Comparison} compares our results to previous works that use one or more of WL, X-ray, and the thermal and kinematic Sunyaev Zeldovich effects \citep[see][for a review]{Carlstrom2002SZReview}, which we abbreviate as tSZ and kSZ, respectively. The $S(k)$ inferred by our \BCEmu analysis is generally stronger than those from existing works. However, the BFG-\BCEmu analysis --- which uses a more consistent cosmology prior (Figure \ref{fig:Testfixcosmo}) --- is in better agreement with existing results. We also show comparisons to the \Bacco analysis of \citet{Arico:2021:Bacco} and \citet{Chen:2023:DESY3}, finding good agreement. All \Bacco results are shallower than BFG-\BCEmu given the narrower priors on astrophysics parameters (Figure \ref{fig:PkSuppresion_Comparison}). The results in Figure \ref{fig:PkSuppresion_Comparison} span a range of datasets and analysis choices, but still corroborate the findings of \citet{Bigwood:2024:BaryonsWLkSZ} by showing a consistent preference for $\approx 25\%$ suppression in the matter power spectrum around $k = 4 h/\mpc$ at $z = 0$.

\section{Intrinsic Alignments}\label{appx:IA}

We list our IA constraints in Figure \ref{fig:IA}. As a reminder to the reader, the joint DECam 13k analysis does not assume all three datasets (DECADE NGC, DECADE SGC, and DES Y3) can be modeled using a shared set of five IA parameters. We instead perform a more conservative analysis and use three sets of IA parameters---one for each of the three data vectors. This results in fifteen total IA parameters. We distinguish parameters between the different sets using the superscripts NGC, SGC, and DES.

In Appendix B1 of \citetalias{paper4}, we found the \decade NGC data had a statistically significant preference for a non-zero TATT amplitude, $a_2 > 0$.
We now briefly reproduce the summary from Section 5.2 of \citetalias{paper4} --- namely, at this time, we do not have a clean explanation for the exact origin of the NGC IA constraint. We performed a variety of empirical tests in \citetalias{paper4} which together indicate that the underlying signal does not come from variations in image quality, nor from a specific subset of the data (\eg galaxies with large angular sizes). Deeper investigations of this signal cannot be made using weak lensing measurements alone, and will require cross-matching with spectroscopic data to isolate the IA signal \citep[\eg][]{Samuroff:2019:IA, Samuroff2022, Georgiou:2025:IA, Siegel:2025:IA}. Note that the final cosmology constraints from the NGC data are insensitive to these oddities in the IA behavior. \citetalias{paper4} discusses all additional tests performed in relation to this result.

The IA constraints from the \decade SGC dataset, on the other hand, are consistent with no IA signal. This is somewhat anticipated as the SGC dataset is less precise than the NGC dataset --- we have 63 million galaxies, compared to the 107 million of the NGC. These constraints from \decade SGC are also consistent with those from DES, and consistent with the no-IA scenario ($a_1 = a_2 = 0$) within 1$\sigma$ to 2 $\sigma$. In all cases, the constraints on $b_{\rm TA}$ are not shown as they are prior-dominated, consistent with results from other analyses \citep[][\citetalias{paper4}]{Secco2021, Amon2021}. 

Figure \ref{fig:IA} also shows the change in IA constraints if we forego scale cuts in the data vector. This nearly doubles (692 to 1200) the number of measurements. The IA constraints, however, are consistent across the two analyses. If we then add baryon suppression to our model, the IA constraints continue to be consistent. This confirms that the baryon suppression presented in this work (\eg Figure \ref{fig:PkSuppression}) is not compensating for some change in the IA parameters.

In summary, Figure \ref{fig:IA} showcases that the IA parameters are well behaved, except for the non-zero $a_2$ amplitude of \decade NGC which we discuss in detail in \citetalias{paper4}. The constraints remain relatively unchanged if we utilize all small-scale measurements as well, and if we add baryon suppression to our modeling pipeline. This showcases the relative robustness of the IA measurements extracted from this work.

\section{Data vectors and best-fit model}\label{appx:DV}

\begin{figure*}
    \centering
    \includegraphics[width=2\columnwidth]{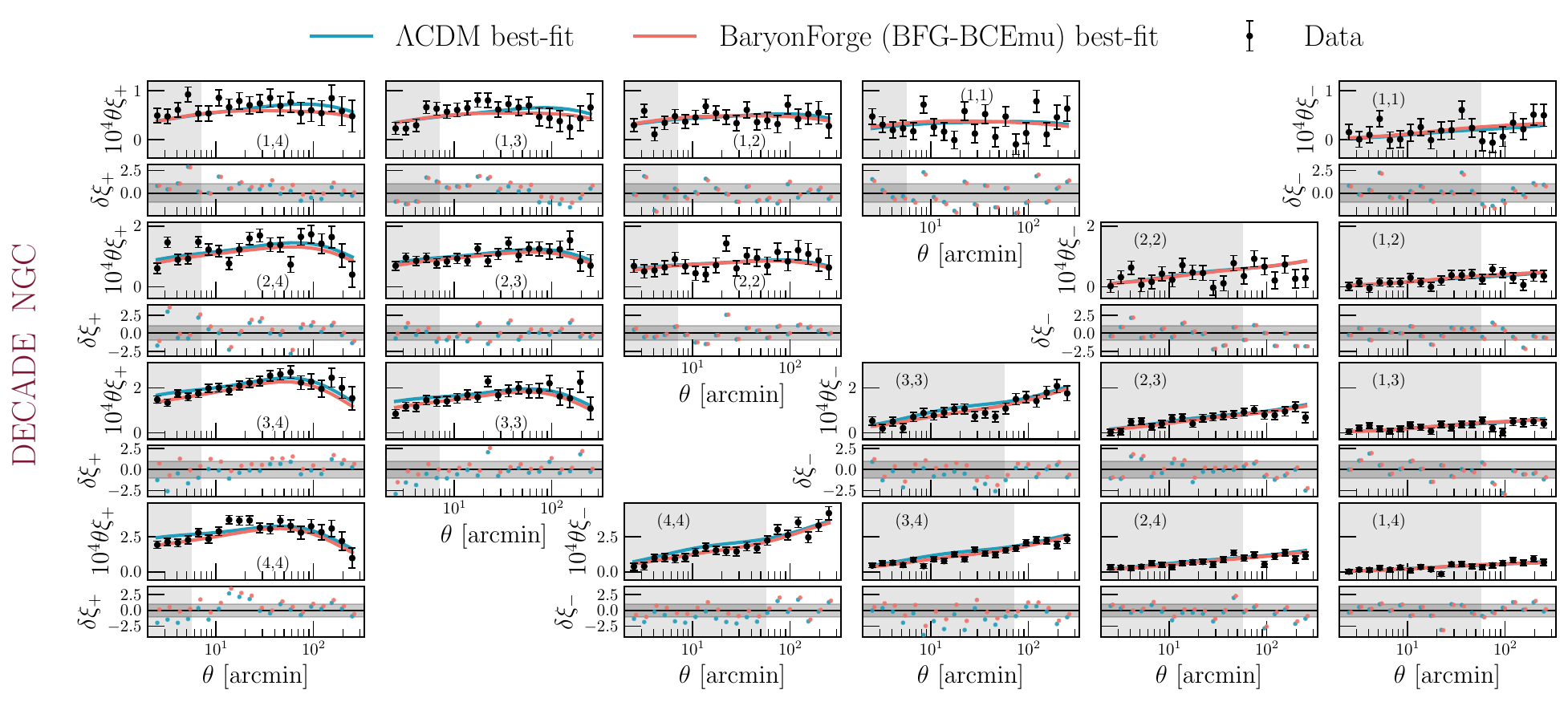}
    \includegraphics[width=2\columnwidth]{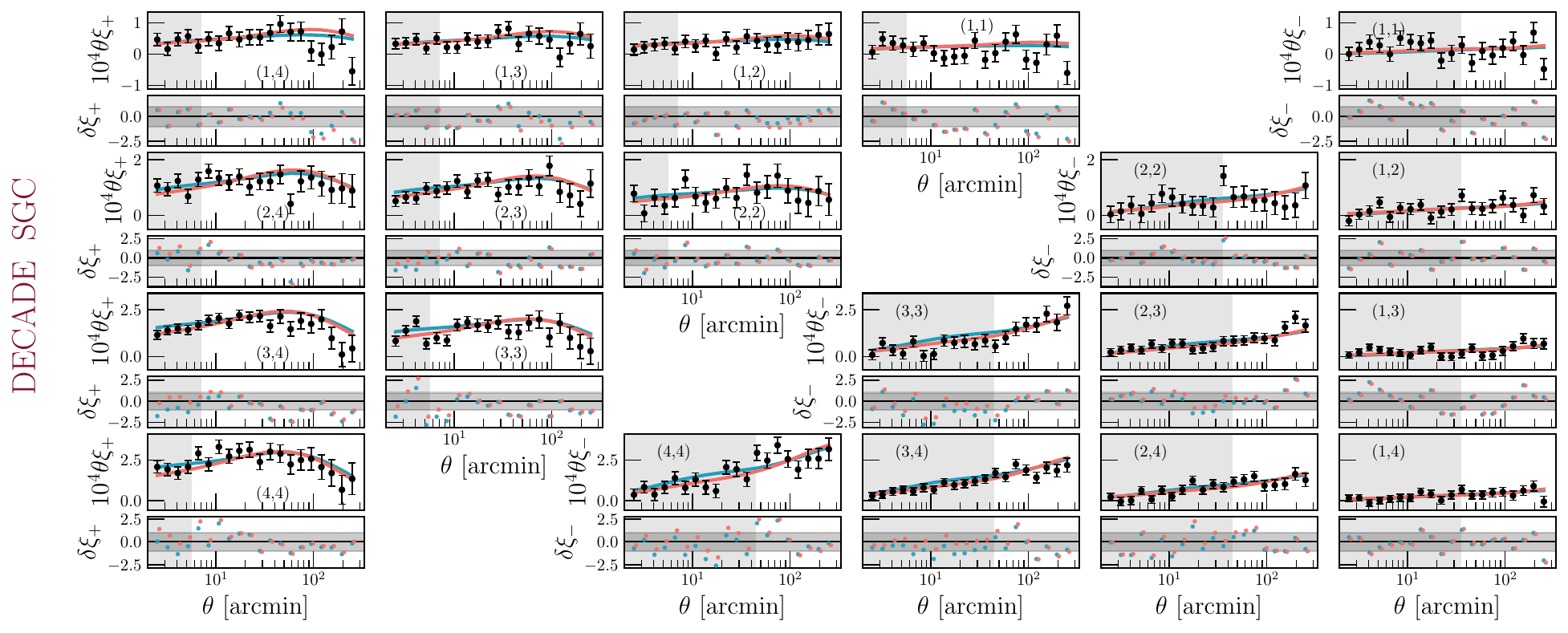}
    \includegraphics[width=2\columnwidth]{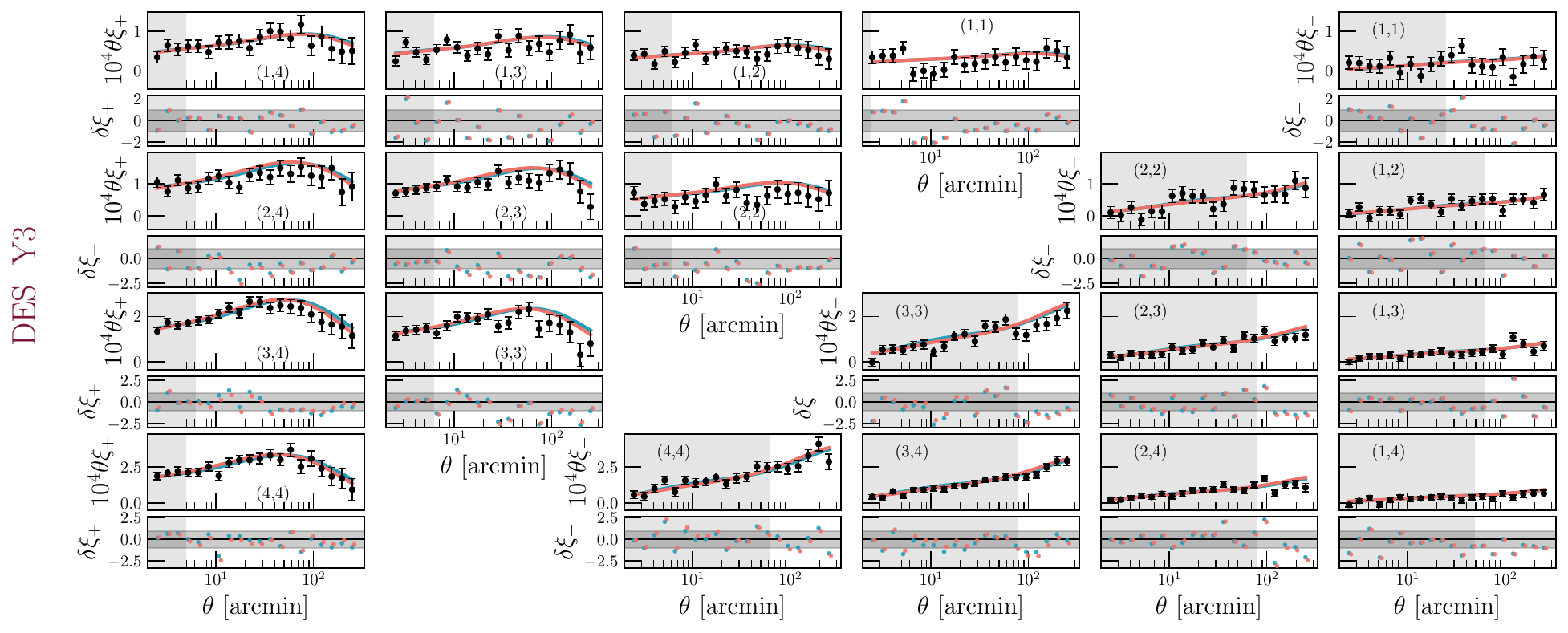}
    \caption{The measured $\xi_\pm$ data vectors for \decade NGC (top), \decade SGC (middle), and for DES Y3 (bottom). We overplot best-fit theory predictions from the Fiducial \LCDM analysis, and from using all scales but including baryon suppression for our small-scale modeling. The gray bands denote the scale-cuts derived from assuming a specific baryon suppression model \citepalias{paper3}.}
    \label{fig:datavectors}
\end{figure*}

Figure \ref{fig:datavectors} shows the cosmic shear data vectors for \decade NGC, \decade SGC, and DES Y3. The scale cuts for each bin are denoted by gray bands. We also overplot the best-fit theory from the fiducial \LCDM analysis, and from the baryon suppression analysis with the \BaryonForge \BCEmu (BFG-\BCEmu) model. The latter uses all available measurements, and not just those unaffected by the marked scale cuts.


\label{lastpage}
\end{document}